\documentclass[aps,prx,twocolumn,secnumarabic,amsmath,amssymb,superscriptaddress]{revtex4-1}
\usepackage{amssymb,amsbsy,graphicx,subfigure,bm,multirow,epstopdf}
\usepackage{color}
\usepackage[colorlinks=true,
linkcolor=blue,
urlcolor=blue,
citecolor=blue]{hyperref}

\begin{document}

\title{Toward Kitaev's sixteenfold way in a honeycomb lattice model}

\author{Shang-Shun~Zhang}

\affiliation{Department of Physics and Astronomy, The University of
Tennessee, Knoxville, Tennessee 37996, USA}

\author{Cristian~D.~Batista}

\affiliation{Department of Physics and Astronomy, The University of
Tennessee, Knoxville, Tennessee 37996, USA}

\affiliation{Neutron Scattering Division and Shull-Wollan Center,
Oak Ridge National Laboratory, Oak Ridge, Tennessee 37831, USA}

\author{G{\'a}bor~B.~Hal{\'a}sz}

\affiliation{Materials Science and Technology Division, Oak Ridge
National Laboratory, Oak Ridge, Tennessee 37831, USA}

\begin{abstract}

Kitaev's sixteenfold way is a classification of exotic topological
orders in which $\mathbb{Z}_2$ gauge theory is coupled to Majorana
fermions of Chern number $C$. The $16$ distinct topological orders
within this class, depending on $C \, \mathrm{mod} \, 16$, possess a
rich variety of Abelian and non-Abelian anyons. We realize more than
half of Kitaev's sixteenfold way, corresponding to Chern numbers
$0$, $\pm 1$, $\pm 2$, $\pm 3$, $\pm 4$, and $\pm 8$, in an exactly
solvable generalization of the Kitaev honeycomb model. For each
topological order, we explicitly identify the anyonic excitations
and confirm their topological properties. In doing so, we observe
that the interplay between lattice symmetry and anyon permutation
symmetry may lead to a ``weak supersymmetry'' in the anyon spectrum.
The topological orders in our honeycomb lattice model could be
directly relevant for honeycomb Kitaev materials, such as
$\alpha$-RuCl$_3$, and would be distinguishable by their specific
quantized values of the thermal Hall conductivity.

\end{abstract}

\maketitle

\section{Introduction} \label{sec-int}

Topological order is an important cornerstone of modern condensed
matter physics which facilitates a classification of gapped phases
of matter beyond the classical paradigm of spontaneous symmetry
breaking \cite{Wen-2004}. While topologically ordered phases may be
fully symmetric and locally featureless, they are characterized by
particular patterns of long-range quantum entanglement
\cite{Chen-2010} which manifest in robust global features, such as a
topological ground-state degeneracy \cite{Wen-1990} and a universal
correction to the bipartite entanglement entropy
\cite{Preskill-2006, Levin-2006}.

Arguably, the most exciting feature of topological order is the
fractionalization of fundamental particles into emergent nonlocal
quasiparticles. Because of their nonlocal nature, these
fractionalized quasiparticles possess unusual ``anyonic'' particle
statistics in two dimensions that is distinct from both bosons and
fermions. In particular, moving one anyon around another one
(``braiding'') may correspond to a nontrivial operation on the
underlying quantum state \cite{Kitaev-2003}. For Abelian topological
orders, these braiding operations act on a single quantum state,
while for non-Abelian topological orders, they act on a set of
degenerate quantum states within an internal space spanned by the
anyons themselves. In addition to their fundamental scientific
appeal, such non-Abelian anyons are highly promising from the
perspective of topological quantum computation \cite{Nayak-2008}.

Each topological order is uniquely characterized by the topological
properties of its anyonic quasiparticle excitations: the distinct
classes of anyons as well as the fusion and braiding rules between
them \cite{Kitaev-2006}. To a large extent, anyons generalize the
concept of topological defects in classically ordered systems
\cite{Mermin-1979}. Indeed, the anyon classes are topologically
distinct in the sense that they cannot be locally transformed into
each other, while the fusion rules between these classes are
analogous to the combination rules between topological defects.
Together with the fundamentally quantum braiding rules, these
topological properties fully define a given topological order,
mathematically described in the language of topological quantum
field theory \cite{Bernevig-2015}.

The simplest and most widely studied topological order is
$\mathbb{Z}_2$ gauge theory \cite{Kitaev-2003}, which gives rise to
an entire class of topological orders when coupled to gapped
Majorana fermions of Chern number $C$ \cite{Kitaev-2006}. This class
contains an infinite number of topologically distinct edge theories
as the number of chiral Majorana edge modes is given by the Majorana
Chern number $C$ itself. Interestingly, however, the bulk
topological order is determined by $C \, \mathrm{mod} \, 16$, and
the infinitely many edge theories thus correspond to only $16$ bulk
topological orders with distinct topological properties of the bulk
anyons.

This classification, commonly known as Kitaev's sixteenfold way
\cite{Kitaev-2006}, contains both Abelian and non-Abelian
topological orders, corresponding to even and odd Majorana Chern
numbers, respectively. The topological orders of Kitaev's
sixteenfold way are relevant for a wide range of topological
materials, including fractional quantum Hall systems, topological
superconductors, as well as quantum spin liquids. In particular,
recent thermal Hall conductivity measurements in the quantum spin
liquid candidate $\alpha$-RuCl$_3$ \cite{Kasahara-2018} indicate a
single Majorana edge mode for a range of applied magnetic fields,
corresponding to the non-Abelian $C = 1$ topological order.

The search for topological orders in such magnetic materials was
fueled by the discovery of the Kitaev honeycomb model
\cite{Kitaev-2006}, which realizes the $C = 0$ and $C = \pm 1$
topological orders in an exactly solvable spin model on the
honeycomb lattice. Indeed, the bond-dependent Ising interactions of
this exactly solvable model were first proposed to emerge between
transition-metal ions in the $d^5$ \cite{Jackeli-2009,
Chaloupka-2010} and $d^7$ \cite{Liu-2018, Sano-2018} configurations
as well as between rare-earth ions \cite{Li-2017, Jang-2019}, and
then these proposals led to a wide range of honeycomb candidate
materials, including (Na,Li)$_2$IrO$_3$ \cite{Singh-2010, Liu-2011,
Singh-2012, Choi-2012, Ye-2012, Comin-2012, Chun-2015,
Williams-2016}, H$_3$LiIr$_2$O$_6$ \cite{Kitagawa-2018},
$\alpha$-RuCl$_3$ \cite{Plumb-2014, Sandilands-2015, Sears-2015,
Majumder-2015, Johnson-2015, Sandilands-2016, Banerjee-2016,
Banerjee-2017, Do-2017}, Na$_3$Co$_2$SbO$_6$ \cite{Yan-2019}, and
YbCl$_3$ \cite{Xing-2019, Sala-2019}. However, it should be
emphasized that, while the original Kitaev model only contains $|C|
\leq 1$ topological orders, there is no reason to believe that only
these topological orders can emerge in such honeycomb magnets.

In this work, we study an exactly solvable generalization
\cite{Zhang-2019} of the Kitaev model that respects all symmetries
of the honeycomb lattice and realizes more than half of the
topological orders in Kitaev's sixteenfold way, corresponding to
Majorana Chern numbers $0$, $\pm 1$, $\pm 2$, $\pm 3$, $\pm 4$, and
$\pm 8$. These topological orders contain both Abelian and
non-Abelian anyons with a rich variety of fusion and braiding rules,
and are experimentally distinguishable by their different quantized
values of the thermal Hall conductivity. For each topological order,
we use the exact solution of our model to explicitly identify the
anyon classes and verify their fusion rules. In some cases, we find
that lattice symmetry becomes intertwined with anyon permutation
symmetry, corresponding to weak symmetry breaking
\cite{Kitaev-2006}, and gives rise to a ``weak supersymmetry'' in
the excitation spectrum. Since the additional four-spin interactions
of our generalized Kitaev model arise naturally from
time-reversal-symmetric perturbations \cite{Zhang-2019}, in the same
way as the three-spin interactions in the original Kitaev model
arise from an external magnetic field, we believe that the $|C| > 1$
topological orders described in this work are likely to be realized
in spin-orbit-coupled honeycomb magnets, such as $\alpha$-RuCl$_3$.

\begin{figure}[tbp]
\centering
\includegraphics[width=1.0\columnwidth]{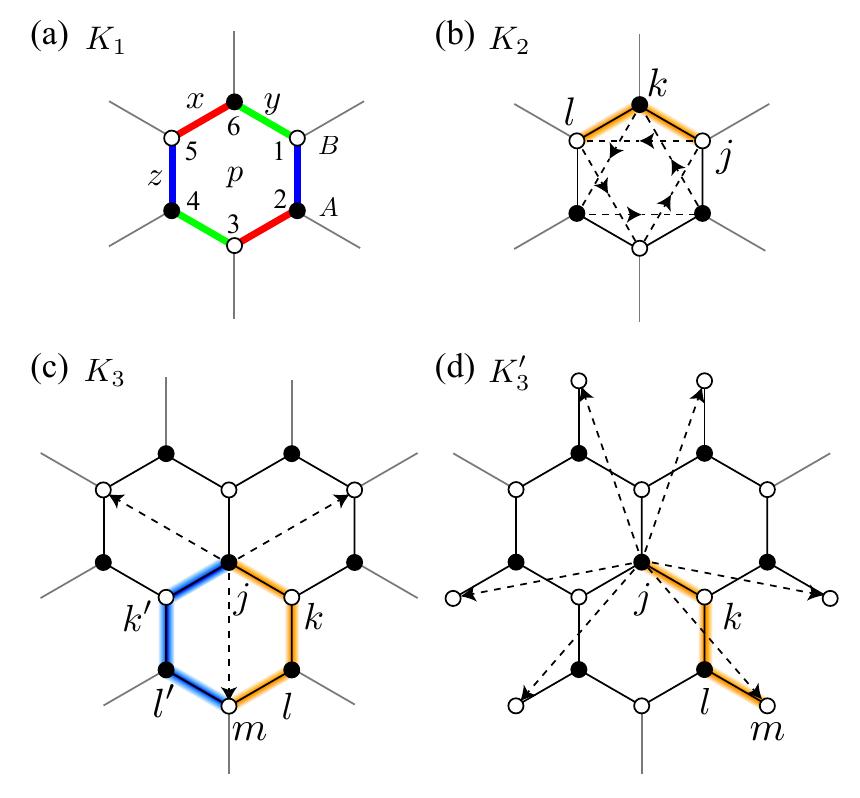}
\caption{Generalized Kitaev model. (a) Bond-dependent Ising
interactions of the $K_1$ term corresponding to the pure Kitaev
model: the spin components $\sigma^{x,y,z}$ at neighboring honeycomb
sites are coupled along $x$ (red), $y$ (green), and $z$ (blue)
bonds, respectively. The site-labeling convention around a plaquette
$p$ is also illustrated. (b) Representative (orange) path $\langle
jkl \rangle_{yx}$ associated with the $K_2$ term in
Eq.~(\ref{eq-mod-H-1-2}). (c)-(d) Representative (orange) paths
$\langle jklm \rangle_{yzx}$ (c) and $\langle jklm \rangle_{yzy}$
(d) associated with the $K_3$ and $K_3'$ terms in
Eq.~(\ref{eq-mod-H-1-3}), respectively. Spin interactions along
these paths give rise to Majorana hopping terms along the dashed
arrows. Note that the $K_3$ interactions come in symmetry-related
pairs (orange and blue) that correspond to the same Majorana hopping
term and may interfere constructively or destructively. In general,
sites in sublattice $A$ ($B$) are marked by black (white) dots.}
\label{fig-1}
\end{figure}

\section{Lattice model} \label{sec-mod}

We consider a generalization of the Kitaev spin model on the
honeycomb lattice,
\begin{equation}
\mathcal{H} = \mathcal{H}_1 + \mathcal{H}_2 + \mathcal{H}_3,
\label{eq-mod-H-1}
\end{equation}
where the first term
\begin{equation}
\mathcal{H}_1 = - K_1 \sum_{\alpha} \sum_{\langle jk
\rangle_{\alpha}} \sigma_j^{\alpha} \sigma_k^{\alpha}
\label{eq-mod-H-1-1}
\end{equation}
is the pure Kitaev model \cite{Kitaev-2006} with Ising interactions
between the spin components $\sigma^{\alpha}$ along each $\alpha =
\{ x,y,z \}$ bond $\langle jk \rangle_{\alpha}$ [see
Fig.~\ref{fig-1}(a)], while the remaining two terms $\mathcal{H}_r$
with $r = 2,3$ contain products of such Ising interactions along
paths consisting of $r$ bonds each. If we define $\langle jkl
\rangle_{\alpha \beta}$ to be the path consisting of the two bonds
$\langle jk \rangle_{\alpha}$ and $\langle kl \rangle_{\beta}$ [see
Fig.~\ref{fig-1}(b)], the second term reads
\begin{eqnarray}
\mathcal{H}_2 &=& -i K_2 \sum_{(\alpha \beta \gamma)} \sum_{\langle
jkl \rangle_{\alpha \beta}} \epsilon_{(\alpha \beta \gamma)} \big(
\sigma_j^{\alpha} \sigma_k^{\alpha} \big) \big( \sigma_k^{\beta}
\sigma_l^{\beta}
\big) \nonumber \\
&=& K_2 \sum_{(\alpha \beta \gamma)} \sum_{\langle jkl
\rangle_{\alpha \beta}} \sigma_j^{\alpha} \sigma_k^{\gamma}
\sigma_l^{\beta}, \label{eq-mod-H-1-2}
\end{eqnarray}
where $(\alpha \beta \gamma)$ is a general permutation of $(xyz)$,
and $\epsilon_{(\alpha \beta \gamma)}$ is $+1$ ($-1$) for even (odd)
permutations. Using analogous notation, the third term then takes
the form
\begin{eqnarray}
\mathcal{H}_3 &=& -K_3 \sum_{(\alpha \beta \gamma)} \sum_{\langle
jklm \rangle_{\alpha \beta \gamma}} \big( \sigma_j^{\alpha}
\sigma_k^{\alpha} \big) \big( \sigma_k^{\beta} \sigma_l^{\beta}
\big) \big( \sigma_l^{\gamma} \sigma_m^{\gamma} \big)
\nonumber \\
&&- K_3' \sum_{(\alpha \beta \gamma)} \sum_{\langle jklm
\rangle_{\alpha \beta \alpha}} \big( \sigma_j^{\alpha}
\sigma_k^{\alpha} \big) \big( \sigma_k^{\beta} \sigma_l^{\beta}
\big) \big( \sigma_l^{\alpha} \sigma_m^{\alpha} \big) \nonumber \\
&=& K_3 \sum_{(\alpha \beta \gamma)} \sum_{\langle jklm
\rangle_{\alpha \beta \gamma}} \sigma_j^{\alpha} \sigma_k^{\gamma}
\sigma_l^{\alpha} \sigma_m^{\gamma}
\label{eq-mod-H-1-3} \\
&&- K_3' \sum_{(\alpha \beta \gamma)} \sum_{\langle jklm
\rangle_{\alpha \beta \alpha}} \sigma_j^{\alpha} \sigma_k^{\gamma}
\sigma_l^{\gamma} \sigma_m^{\alpha}, \nonumber
\end{eqnarray}
where $\langle jklm \rangle_{\alpha \beta \gamma}$ and $\langle jklm
\rangle_{\alpha \beta \alpha}$ are paths consisting of three bonds
each [see Figs.~\ref{fig-1}(c) and \ref{fig-1}(d)]. As it is clear
from our construction, the term $\mathcal{H}_r$ for general $r$
contains $(r+1)$-spin interactions and thus breaks (preserves)
time-reversal symmetry for even (odd) $r$. We remark that the term
$\mathcal{H}_2$ was already introduced in Ref.~\cite{Kitaev-2006}
while the term $\mathcal{H}_3$ was first considered in
Ref.~\cite{Zhang-2019}. It is also important to note that these two
terms are respectively generated by time-reversal-breaking and
time-reversal-symmetric perturbations on top of the pure Kitaev
model.

Remarkably, the generalized Kitaev model in Eq.~(\ref{eq-mod-H-1})
is exactly solvable in the same way as the original Kitaev model
\cite{Kitaev-2006}. By expressing each physical spin component as a
product of two Majorana fermions, $\sigma_j^{\alpha} = i
b^{\alpha}_j c_j^{\phantom{\dag}}$, the Hamiltonians $\mathcal{H}_n$
in Eqs.~(\ref{eq-mod-H-1-1})-(\ref{eq-mod-H-1-3}) become
\begin{eqnarray}
\mathcal{H}_1 &=& i K_1 \sum_{\alpha} \sum_{\langle jk
\rangle_{\alpha}} u_{jk}^{\alpha} c_j^{\phantom{\dag}}
c_k^{\phantom{\dag}}, \nonumber \\
\mathcal{H}_2 &=& -i K_2 \sum_{(\alpha \beta \gamma)} \sum_{\langle
jkl \rangle_{\alpha \beta}} \epsilon_{(\alpha \beta \gamma)}
u_{jk}^{\alpha} u_{lk}^{\beta} c_j^{\phantom{\dag}}
c_l^{\phantom{\dag}}, \nonumber \\
\mathcal{H}_3 &=& i K_3 \sum_{(\alpha \beta \gamma)} \sum_{\langle
jklm \rangle_{\alpha \beta \gamma}} u_{jk}^{\alpha} u_{lk}^{\beta}
u_{lm}^{\gamma} c_j^{\phantom{\dag}}
c_m^{\phantom{\dag}} \label{eq-mod-H-2} \\
&& +i K_3' \sum_{(\alpha \beta \gamma)} \sum_{\langle jklm
\rangle_{\alpha \beta \alpha}} u_{jk}^{\alpha} u_{lk}^{\beta}
u_{lm}^{\alpha} c_j^{\phantom{\dag}} c_m^{\phantom{\dag}}, \nonumber
\end{eqnarray}
where the $\mathbb{Z}_2$ gauge fields $u_{jk}^{\alpha} =
-u_{kj}^{\alpha} \equiv i b_j^{\alpha} b_k^{\alpha}$ along the bonds
$\langle jk \rangle_{\alpha}$ are conserved quantities that commute
with each other. Therefore, the Hamiltonian $\mathcal{H}$ in
Eq.~(\ref{eq-mod-H-1}) describes free fermions coupled to a static
$\mathbb{Z}_2$ gauge theory, and $c_j$ can be identified as
deconfined Majorana fermion (``spinon'') degrees of freedom. In
terms of these Majorana fermions, each term $\mathcal{H}_r$ in
Eq.~(\ref{eq-mod-H-2}) corresponds to $r$-th-neighbor hopping
\cite{footnote-1}. Also, unlike the gauge fields themselves, the
product of the gauge fields around any plaquette $p$ [see
Fig.~\ref{fig-1}(a)] is a gauge-invariant quantity that can be
expressed in terms of the physical spins:
\begin{equation}
W_p = u_{12}^z u_{32}^x u_{34}^y u_{54}^z u_{56}^x u_{16}^y =
\sigma_1^x \sigma_2^y \sigma_3^z \sigma_4^x \sigma_5^y \sigma_6^z.
\label{eq-mod-W}
\end{equation}
Thus, $W_p = \pm 1$ can be identified as static $\mathbb{Z}_2$ gauge
flux (``vison'') degrees of freedom.

\begin{figure}[tbp]
\centering
\includegraphics[width=1.0\columnwidth]{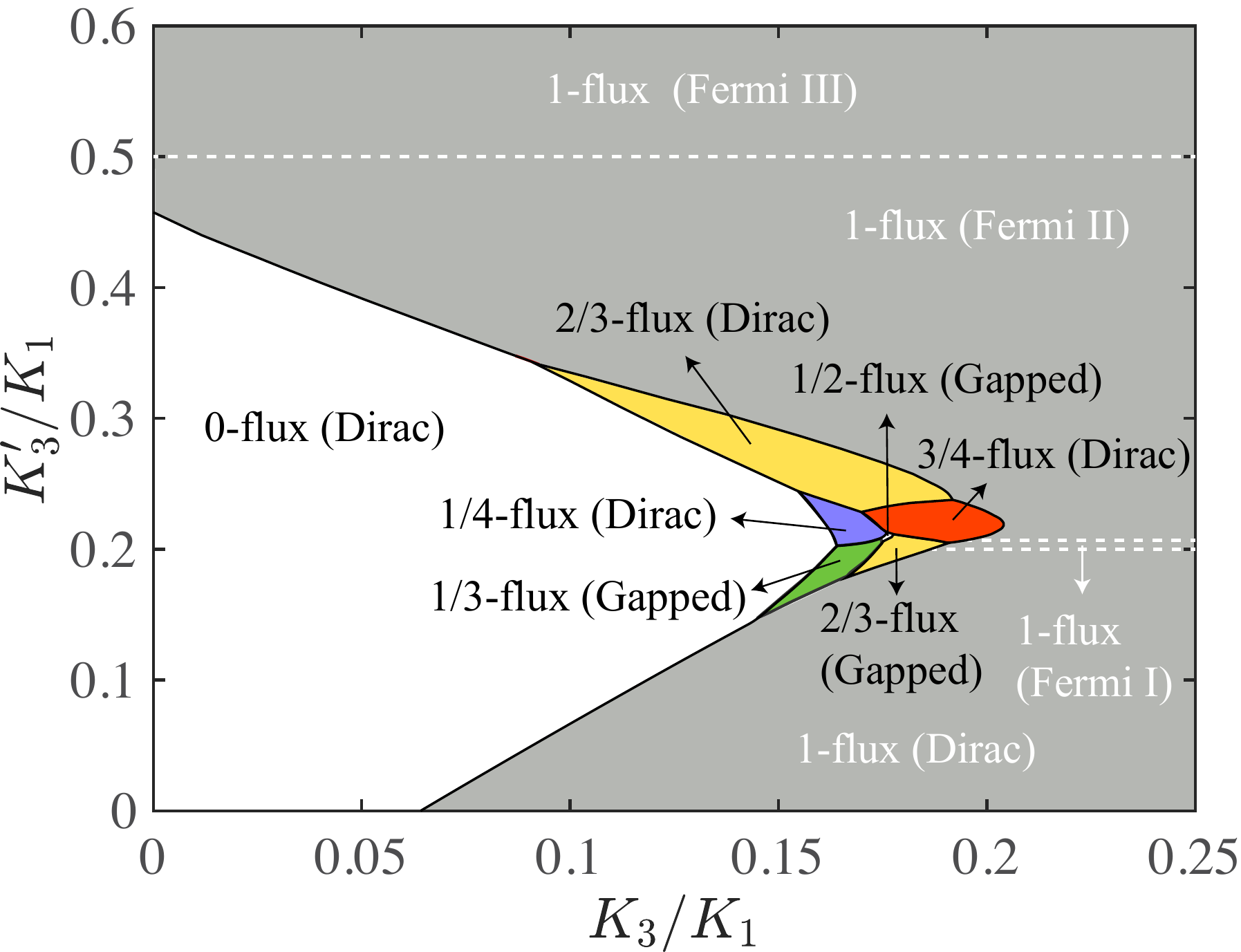}
\caption{Phase diagram of the time-reversal-symmetric Hamiltonian
$\mathcal{H}_1 + \mathcal{H}_3$ as a function of $K_3 / K_1$ and
$K_3' / K_1$ \cite{Zhang-2019}. Black solid lines are first-order
transitions between different flux sectors, denoted by distinct
colors, while white dashed lines are second-order Lifshitz
transitions between different Majorana nodal structures, specified
in parentheses.} \label{fig-2}
\end{figure}

\begin{figure*}[tbp]
\centering
\includegraphics[width=1.3\columnwidth]{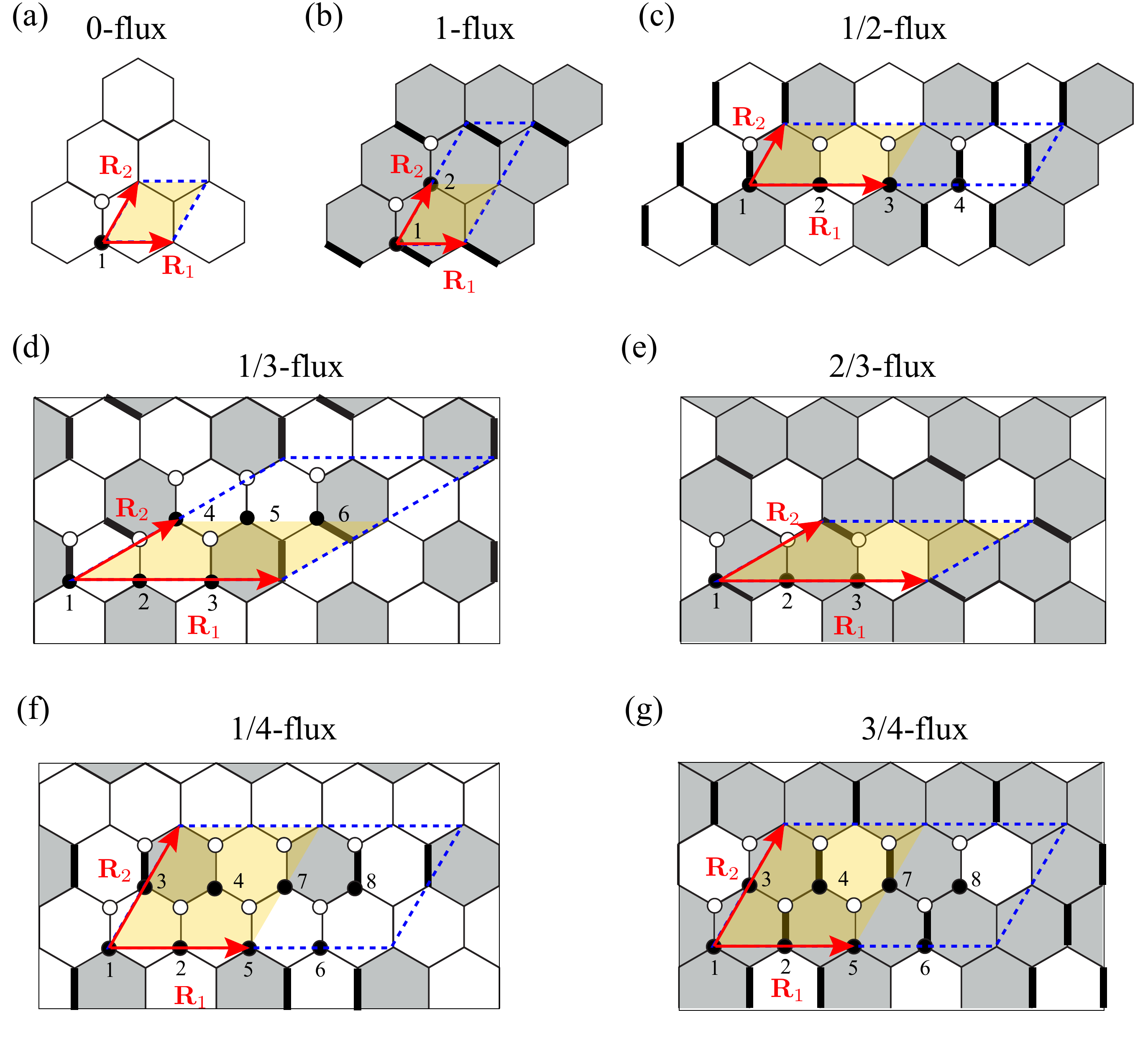}
\caption{Flux configurations and representative gauge-field
configurations in the ground-state flux sectors of Fig.~\ref{fig-2}.
In each case, plaquettes with $W_p = +1$ ($W_p = -1$) are marked by
white (gray) filling, while bonds with $u_{jk}^{\alpha} = +1$
($u_{jk}^{\alpha} = -1$) are marked by thin (thick) lines. The
physical unit cell, spanned by the lattice vectors
$\mathbf{R}_{1,2}$, is marked by a yellow shaded parallelogram,
while the Majorana unit cell is marked by a blue dashed
parallelogram. Note that the Majorana unit cell may contain several
honeycomb unit cells, indexed by $\nu = 1,\dots, n$, each containing
one $A$ site (black dot) and one $B$ site (white dot).}
\label{fig-3}
\end{figure*}

While Eq.~(\ref{eq-mod-H-2}) reduces to a quadratic fermion problem
in each flux sector, $\{ W_p = \pm 1 \}$, represented with an
appropriate gauge-field configuration, $\{ u_{jk}^{\alpha} = \pm 1
\}$, it is not immediately clear which flux sector contains the
ground state of the physical spin model $\mathcal{H}$. For the pure
Kitaev model $\mathcal{H}_1$, it is guaranteed by Lieb's theorem
\cite{Lieb-1994} that the ground state belongs to the $0$-flux
sector characterized by $W_p = +1$ for all $p$. However, Lieb's
theorem no longer applies if the additional terms $\mathcal{H}_2$
and/or $\mathcal{H}_3$ are included in the spin model. Indeed, it
was demonstrated in Ref.~\cite{Zhang-2019} that the frustration
between $\mathcal{H}_1$ and $\mathcal{H}_3$ can stabilize a wide
range of flux sectors as a function of $K_3 / K_1$ and $K_3' / K_1$
(see Fig.~\ref{fig-2}), including the $1$-flux sector characterized
by $W_p = -1$ for all $p$, as well as fractional-flux sectors in
which a nontrivial fraction of the plaquettes have $W_p = -1$ rather
than $W_p = +1$. In these fractional-flux sectors, the plaquettes
with $W_p = -1$ form crystalline structures (``vison crystals'')
that spontaneously break translation symmetry (see
Fig.~\ref{fig-3}).

Assuming an infinitesimally small coupling constant $K_2$, we start
from the time-reversal-symmetric Hamiltonian $\mathcal{H}_1 +
\mathcal{H}_3$ \cite{Zhang-2019} and treat the Hamiltonian term
$\mathcal{H}_2$ as a time-reversal-breaking perturbation. Due to the
finite flux gap, the ground-state flux sectors in Fig.~\ref{fig-2}
are robust against small perturbations. In contrast, if the Majorana
fermions are originally gapless, even an infinitesimally small
time-reversal-breaking perturbation can have a dramatic effect on
their low-energy physics \cite{Kitaev-2006}.

\section{Majorana problems} \label{sec-maj}

\subsection{Quadratic Hamiltonians} \label{sec-ham}

For each ground-state flux sector in Fig.~\ref{fig-2}, the
gauge-field configuration in Fig.~\ref{fig-3} gives rise to a
quadratic Majorana problem [see Eq.~(\ref{eq-mod-H-2})]. The unit
cell of this Majorana problem may consist of $n > 1$ honeycomb unit
cells for two distinct reasons. First, the physical unit cell is
enlarged in the fractional-flux sectors because translation symmetry
is spontaneously broken. This enlargement is twofold for the
$1/2$-flux sector, threefold for the $1/3$-flux and $2/3$-flux
sectors, and fourfold for the $1/4$-flux and $3/4$-flux sectors.
Second, if the physical unit cell has an odd number of $W_p = -1$
plaquettes, translation symmetry acts projectively on the Majorana
fermions. In this case, the Majorana unit cell, as characterized by
the gauge-field configuration, must consist of two physical unit
cells. Consequently, the Majorana unit cell has an additional
twofold enlargement in all flux sectors except for the $0$-flux and
$2/3$-flux sectors.

For each flux sector, we label the honeycomb sites as $j =
(\mathbf{r}, \lambda)$ and the corresponding Majorana fermions as
$c_j = c_{\mathbf{r}, \lambda}$, where $\mathbf{r}$ is the lattice
vector of the Majorana unit cell, and $\lambda = (\mu, \nu)$ in
terms of the sublattice index $\mu = A,B$ and the index $\nu = 1,
\ldots, n$ specifying the particular honeycomb unit cell within the
Majorana unit cell. Using this labeling convention, the quadratic
Majorana Hamiltonian takes the general form
\begin{equation}
\mathcal{H} = \frac{i}{2} \sum_{\mathbf{r}, \mathbf{r}'}
\sum_{\lambda, \lambda'} \tilde{H}_{\mathbf{r}' - \mathbf{r},
\lambda, \lambda'}^{\phantom{\dag}} c_{\mathbf{r},
\lambda}^{\phantom{\dag}} c_{\mathbf{r'},
\lambda'}^{\phantom{\dag}}, \label{eq-ham-H-1}
\end{equation}
where each $\tilde{H}_{\mathbf{r}' - \mathbf{r}, \lambda, \lambda'}$
is proportional to the product of the static gauge fields
$u_{jk}^{\alpha} = \pm 1$ along a path connecting the sites
$(\mathbf{r}, \lambda)$ and $(\mathbf{r}', \lambda')$. Introducing
the momentum-space complex fermions
\begin{equation}
\psi_{\mathbf{q}, \lambda}^{\phantom{\dag}} = \frac{1} {\sqrt{N}}
\sum_{\mathbf{r}} c_{\mathbf{r}, \lambda}^{\phantom{\dag}} e^{-i
\mathbf{q} \cdot \mathbf{r}}, \label{eq-ham-psi-1}
\end{equation}
where $N$ is the number of honeycomb sites, and arranging them into
the $2n$-component vector
\begin{equation}
\psi_{\mathbf{q}}^{\phantom{\dag}} \equiv \left[ \psi_{\mathbf{q},
(A,1)}^{\phantom{\dag}}, \ldots, \psi_{\mathbf{q},
(A,n)}^{\phantom{\dag}}, \psi_{\mathbf{q}, (B,1)}^{\phantom{\dag}},
\ldots, \psi_{\mathbf{q}, (B,n)}^{\phantom{\dag}} \right]^T,
\label{eq-ham-psi-2}
\end{equation}
the Hamiltonian in Eq.~(\ref{eq-ham-H-1}) can then be written as
\begin{equation}
\mathcal{H} = \sum_{\mathbf{q}} \psi_{\mathbf{q}}^{\dag} \cdot
H_{\mathbf{q}}^{\phantom{\dag}} \cdot
\psi_{\mathbf{q}}^{\phantom{\dag}}, \label{eq-ham-H-2}
\end{equation}
where $H_{\mathbf{q}}$ is a $2n \times 2n$ matrix with elements
\begin{equation}
\left( H_{\mathbf{q}} \right)_{\lambda \lambda'} = \sum_{\mathbf{r}}
i \tilde{H}_{\mathbf{r}, \lambda, \lambda'} e^{i \mathbf{q} \cdot
\mathbf{r}}. \label{eq-ham-H-3}
\end{equation}
By diagonalizing the matrix $H_{\mathbf{q}}$ at each momentum
$\mathbf{q}$, one obtains Majorana bands at both positive and
negative energies. However, since $\psi_{-\mathbf{q},
\lambda}^{\phantom{\dag}} = \psi_{\mathbf{q}, \lambda}^{\dag}$ by
definition, there is a redundancy in our description, and only the
bands with positive energies are physical.

\subsection{Projective symmetries} \label{sec-sym}

We now discuss the general symmetries of the Majorana problem in
Eq.~(\ref{eq-ham-H-2}). Since the Majorana fermions are
fractionalized degrees of freedom, symmetries may act on them
projectively \cite{Wen-2002}, i.e., the classical relations between
symmetry operations may only be satisfied up to an overall complex
phase factor $e^{i \varphi}$. However, as the Majorana fermions are
coupled to $\mathbb{Z}_2$ gauge fields, this phase factor must
actually be a sign $\pm 1$ \cite{You-2012}.

We first consider the translation symmetries $\mathcal{T}_1$ and
$\mathcal{T}_2$ along the lattice vectors $\mathbf{R}_1$ and
$\mathbf{R}_2$ of the physical unit cell (see Fig.~\ref{fig-3}).
Note that the physical unit cell depends on the particular flux
sector and may be larger than the original honeycomb unit cell due
to spontaneous breaking of translation symmetry in the
fractional-flux sectors. If the physical unit cell has no overall
$\mathbb{Z}_2$ flux, corresponding to an even number of $W_p = -1$
plaquettes, translation symmetry acts linearly (i.e., not
projectively) on the Majorana fermions, and the two elementary
translations commute: $[\mathcal{T}_1, \mathcal{T}_2] = 0$. The
Brillouin zone is then spanned by the reciprocal lattice vectors
$\mathbf{G}_1$ and $\mathbf{G}_2$ corresponding to the physical unit
cell, and different points in the Brillouin zone are labeled by
different eigenvalues of the translations $\mathcal{T}_1$ and
$\mathcal{T}_2$.

\begin{figure}[tbp]
\centering
\includegraphics[width=1.0\columnwidth]{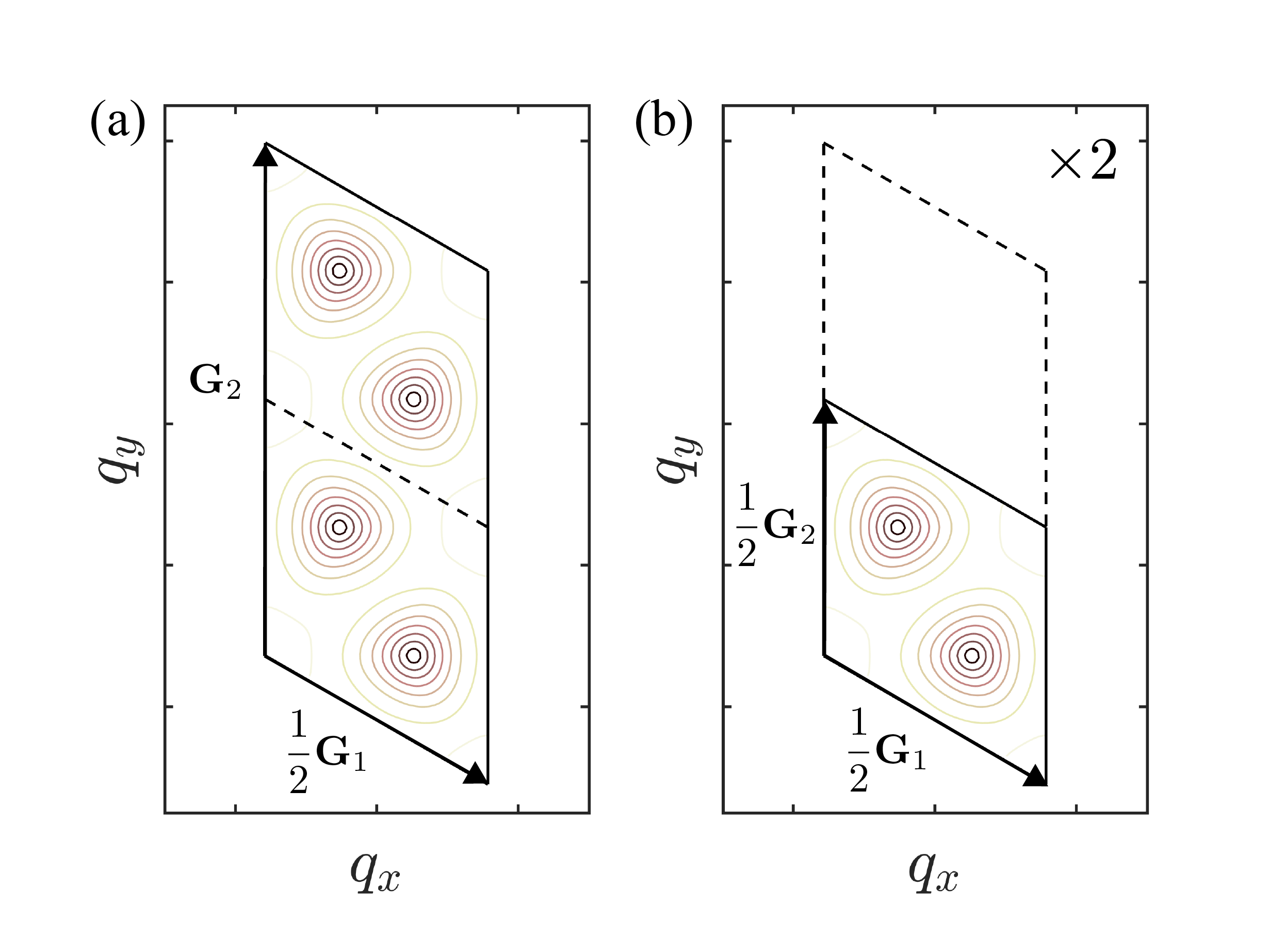}
\caption{Schematic illustrations of the conventional (a) and the
compact (b) Brillouin zones when translation symmetry acts
projectively on the Majorana fermions. The conventional Brillouin
zone is equivalent to two identical copies (``$\times 2$'') of the
compact Brillouin zone.} \label{fig-4}
\end{figure}

Conversely, if the physical unit cell has an overall $\mathbb{Z}_2$
flux, corresponding to an odd number of $W_p = -1$ plaquettes,
translation symmetry acts projectively on the Majorana fermions, and
the two elementary translations anticommute: $\{ \mathcal{T}_1,
\mathcal{T}_2 \} = 0$. The eigenvalues of the translations
$\mathcal{T}_1$ and $\mathcal{T}_2$ are then no longer compatible
quantum numbers for the Majorana fermions. Nevertheless, since
$[\mathcal{T}_1^2, \mathcal{T}_2] = 0$, one may consider a larger
Majorana unit cell spanned by $2 \mathbf{R}_1$ and $\mathbf{R}_2$,
which translates into a smaller Brillouin zone spanned by
$\frac{1}{2} \mathbf{G}_1$ and $\mathbf{G}_2$. In fact, the Majorana
spectrum is periodic with respect to an even smaller Brillouin zone
spanned by $\frac{1}{2} \mathbf{G}_1$ and $\frac{1}{2} \mathbf{G}_2$
[see Fig.~\ref{fig-4}(a)] because the residual symmetry
$\mathcal{T}_1$ anticommutes with $\mathcal{T}_2$ and hence
corresponds to a shift $\frac{1}{2} \mathbf{G}_2$ in the Majorana
momentum. Thus, one may use a compact Brillouin zone spanned by
$\frac{1}{2} \mathbf{G}_{1,2}$ and indicate that each Majorana band
has a twofold ``translation degeneracy'' [see Fig.~\ref{fig-4}(b)].
Different points in this compact Brillouin zone are labeled by
$\mathcal{T}_1^2$ and $\mathcal{T}_2^2$, while the two degenerate
Majorana fermions at a given point are labeled by $\mathcal{T}_1$
and mapped onto each other by $\mathcal{T}_2$ (or vice versa).

We next consider time-reversal symmetry $\mathcal{T}$ and inversion
symmetry $\mathcal{P}$. Time reversal is an antiunitary operation,
$\{ \mathcal{T}, i \} = 0$, and is only a symmetry for $K_2 = 0$. In
each flux sector, it acts on the Majorana fermions as
\begin{equation}
\mathcal{T}: \quad c_{\mathbf{r}, (A, \nu)} \rightarrow
c_{\mathbf{r}, (A, \nu)}, \quad c_{\mathbf{r}, (B, \nu)} \rightarrow
-c_{\mathbf{r}, (B, \nu)}, \label{eq-sym-T}
\end{equation}
and hence satisfies $\mathcal{T}^2 = +1$. In contrast to time
reversal, inversion is a unitary operation, $[\mathcal{P}, i] = 0$,
and is a general symmetry of our model. While the action of
inversion on the Majorana fermions depends on the given flux sector,
it always exchanges the two sublattices $A$ and $B$, and thus
necessarily anticommutes with time reversal: $\{ \mathcal{P},
\mathcal{T} \} = 0$. Also, inversion satisfies $\mathcal{P}^2 = -1$
\cite{You-2012} in all flux sectors except for the $3/4$-flux
sector. For simplicity, we ignore the $3/4$-flux sector in the rest
of this work and only return to it briefly in Sec.~\ref{sec-dis}.

Finally, the redundancy in our description, corresponding to
$H_{\mathbf{-q}}^{\phantom{*}} = -H_{\mathbf{q}}^{*}$ [see
Eq.~(\ref{eq-ham-H-3})], gives rise to an emergent antiunitary
particle-hole symmetry $\mathcal{C}$, which satisfies $[\mathcal{C},
\mathcal{T}] = 0$, $[\mathcal{C}, \mathcal{P}] = 0$, and
$\mathcal{C}^2 = +1$. We emphasize that particle-hole symmetry is
actually an antisymmetry as it anticommutes with the Hamiltonian.
While $\mathcal{T}$, $\mathcal{P}$, and $\mathcal{C}$ each reverse
the fermion momentum $\mathbf{q}$, their two independent products
$\mathcal{S} = \mathcal{T} \mathcal{C}$ and $\mathcal{R} =
\mathcal{P} \mathcal{C}$ transform the fermions at momentum
$\mathbf{q}$ among each other:
\begin{eqnarray}
\mathcal{S}: &\quad& \psi_{\mathbf{q}} \rightarrow S \cdot
\psi_{\mathbf{q}}, \nonumber \\
\mathcal{R}: &\quad& \psi_{\mathbf{q}} \rightarrow R \cdot
\psi_{\mathbf{q}}. \label{eq-sym-SR-1}
\end{eqnarray}
The $2n \times 2n$ transformation matrices are given by
\begin{equation}
S = \left( \begin{array}{cc} I & 0 \\ 0 & -I \end{array} \right),
\quad \,\, R = \left( \begin{array}{cc} 0 & P \\ -P^T & 0
\end{array} \right) \mathcal{K}, \label{eq-sym-SR-2}
\end{equation}
where $\mathcal{K}$ denotes complex conjugation, $I$ is the $n
\times n$ unit matrix, and $P$ is an $n \times n$ permutation matrix
satisfying $P \cdot P^T = I$. The unitary antisymmetry $\mathcal{S}$
can thus be identified as sublattice symmetry, while the antiunitary
antisymmetry $\mathcal{R}$ can be interpreted as an effective
momentum-conserving particle-hole symmetry.

Since $\mathcal{R}$ is a general antisymmetry of our model, the
Hamiltonian matrix $H_{\mathbf{q}}$ in Eq.~(\ref{eq-ham-H-3})
satisfies $\{ R, H_{\mathbf{q}} \} = 0$, which implies that the
Majorana spectrum is symmetric around zero energy at each momentum
$\mathbf{q}$. Furthermore, in the time-reversal-symmetric limit of
$K_2 = 0$, the antisymmetry $\mathcal{S}$ requires $\{ S,
H_{\mathbf{q}} \} = 0$ and therefore constrains the Hamiltonian
matrix to the form
\begin{equation}
H_{\mathbf{q}}^{\phantom{\dag}} = \left( \begin{array}{cc} 0 &
M_{\mathbf{q}}^{\phantom{\dag}} \\ M_{\mathbf{q}}^{\dag} & 0
\end{array} \right). \label{eq-sym-H}
\end{equation}
In this limit, the eigendecomposition of the $2n \times 2n$ matrix
$H_{\mathbf{q}}$ is equivalent to the singular value decomposition
of the $n \times n$ matrix $M_{\mathbf{q}}$.

\subsection{Generic Majorana nodes} \label{sec-nod}

In terms of the low-energy physics, the gapless nodes of the
momentum-space Majorana spectrum are of particular interest. Because
of the antisymmetry $\mathcal{R}$, a generic nodal momentum
$\mathbf{Q}$ has two zero-energy fermions that correspond to
distinct Majorana bands of opposite energies. By projecting onto
these two low-energy Majorana bands around $\mathbf{q} =
\mathbf{Q}$, one then obtains an effective low-energy theory of the
given Majorana node.

For simplicity, we start our discussion from the
time-reversal-symmetric limit of $K_2 = 0$. Since the Hamiltonian
matrix $H_{\mathbf{Q}}$ takes the form of Eq.~(\ref{eq-sym-H}), we
can choose the low-energy subspace to be spanned by two fermions
located on the two respective sublattices $A$ and $B$,
\begin{eqnarray}
\psi_{\mathbf{q}}^{(1)} &=& \sum_{\nu} \big( u_{\mathbf{Q}}^{*}
\big)_{\nu} \psi_{\mathbf{q}, (A, \nu)}^{\phantom{\dag}},
\nonumber \\
\psi_{\mathbf{q}}^{(2)} &=& \sum_{\nu} \big(
v_{\mathbf{Q}}^{\phantom{*}} \big)_{\nu} \psi_{\mathbf{q}, (B,
\nu)}^{\phantom{\dag}}, \label{eq-nod-psi}
\end{eqnarray}
where $u_{\mathbf{Q}}$ ($v_{\mathbf{Q}}$) is the left (right)
eigenvector of the matrix $M_{\mathbf{Q}}$ corresponding to zero
eigenvalue \cite{footnote-2}. If we project onto these two
low-energy fermions, the antisymmetries $\mathcal{S}$ and
$\mathcal{R}$ are represented with the $2 \times 2$ matrices
\begin{equation}
\hat{S} = \tau_3, \quad \,\, \hat{R} = i \tau_2 \mathcal{K},
\label{eq-nod-SR}
\end{equation}
and the most general Hamiltonian matrix anticommuting with both
$\hat{S}$ and $\hat{R}$ takes the form
\begin{equation}
\hat{H}_{\mathbf{q}} = \beta_1 (\mathbf{q}) \tau_1 + \beta_2
(\mathbf{q}) \tau_2, \label{eq-nod-H-1}
\end{equation}
where $\tau_{1,2,3}$ are the Pauli matrices. Since there are two
independent real coefficients, $\beta_1 (\mathbf{q})$ and $\beta_2
(\mathbf{q})$, that must vanish at the nodal momentum $\mathbf{Q}$
itself, the generic nodal structures in two dimensions are point
nodes. Expanding $\beta_{1,2} (\mathbf{q})$ up to linear order in
$\delta \mathbf{q} \equiv \mathbf{q} - \mathbf{Q} = (\delta q_x,
\delta q_y)$,
\begin{eqnarray}
\beta_1 (\mathbf{q}) &=& \boldsymbol{\gamma}_1 \cdot \delta
\mathbf{q} = \gamma_{1,x} \delta q_x + \gamma_{1,y} \delta q_y,
\nonumber \\
\beta_2 (\mathbf{q}) &=& \boldsymbol{\gamma}_2 \cdot \delta
\mathbf{q} = \gamma_{2,x} \delta q_x + \gamma_{2,y} \delta q_y,
\label{eq-nod-beta}
\end{eqnarray}
these point nodes are generically Dirac nodes with linear
dispersions \cite{footnote-3}. Also, by considering the complex
phase of $\beta_1 (\mathbf{q}) + i \beta_2 (\mathbf{q})$ at
$\mathbf{q} = \mathbf{Q} + (\cos \vartheta, \sin \vartheta) \delta
q$ as a function of $\vartheta$, one can assign a winding number
$W_{\mathbf{Q}}$ to each Dirac node, which is generically given by
\begin{equation}
W_{\mathbf{Q}} = \mathrm{sgn} \det \left( \begin{array}{cc}
\gamma_{1,x} & \gamma_{1,y} \\ \gamma_{2,x} & \gamma_{2,y}
\end{array} \right) = \pm 1 \label{eq-nod-W}
\end{equation}
with $\mathrm{sgn} \, x \equiv x / |x|$. Therefore, each Dirac node
is a stable U(1) vortex protected by time-reversal symmetry or,
equivalently, by sublattice symmetry.

If we then break time-reversal symmetry with an infinitesimally
small $K_2 \neq 0$, the Hamiltonian matrix $\hat{H}_{\mathbf{q}}$
still anticommutes with $\hat{R}$ but no longer with $\hat{S}$.
Thus, its most general form reads
\begin{equation}
\hat{H}_{\mathbf{q}} = \beta_1 (\mathbf{q}) \tau_1 + \beta_2
(\mathbf{q}) \tau_2 + \beta_3 (\mathbf{q}) \tau_3,
\label{eq-nod-H-2}
\end{equation}
where the third coefficient may be expanded up to linear order in
$K_2$ such that $\beta_3 (\mathbf{q}) = m_{\mathbf{q}} K_2$. Since
there are three independent real coefficients, nodes can only emerge
as a result of fine tuning, and the spectrum is generically gapped.
In particular, at each Dirac node of the $K_2 = 0$ limit, the
Hamiltonian matrix becomes
\begin{equation}
\hat{H}_{\mathbf{Q}} = m_{\mathbf{Q}} K_2 \tau_3, \label{eq-nod-H-3}
\end{equation}
and the Dirac node at momentum $\mathbf{Q}$ is thus gapped out by a
fermion mass term $\propto m_{\mathbf{Q}} K_2$.

\section{Majorana Chern numbers} \label{sec-che}

\subsection{Definition and numerical results} \label{sec-num}

Since the Majorana spectrum is symmetric around zero energy at each
momentum $\mathbf{q}$ and generically gapped for $K_2 \neq 0$ in
each ground-state flux sector, diagonalizing the quadratic
Hamiltonian in Eq.~(\ref{eq-ham-H-2}) gives equal numbers of
Majorana bands at strictly positive and strictly negative energies.
Therefore, one can define a ground-state Chern number of the
Majorana fermions by summing the Chern numbers of all the
negative-energy Majorana bands. As it was argued in
Ref.~\cite{Kitaev-2006}, the low-energy physics of the corresponding
topological order is completely determined by this Majorana Chern
number $C$.

Mathematically, the eigendecomposition of the $2n \times 2n$ matrix
$H_{\mathbf{q}}$ in Eq.~(\ref{eq-ham-H-3}) gives $2n$ eigenvalues
$\varepsilon_{\mathbf{q}, \kappa}$ and $2n$ corresponding
eigenvectors $w_{\mathbf{q}, \kappa}$ with $\kappa = 1, \ldots, 2n$
at each Majorana momentum $\mathbf{q}$. The Chern number of each
Majorana band, labeled with $\kappa$, is then obtained as
\begin{equation}
C_{\kappa} = \frac{1} {2\pi} \int_{\textrm{BZ}} d \mathbf{q} \,
F_{\mathbf{q}, \kappa}, \label{eq-num-C-1}
\end{equation}
where the Berry curvature at momentum $\mathbf{q}$ is given by
\begin{equation}
F_{\mathbf{q}, \kappa} = \nabla_{\mathbf{q}} \times
\mathbf{A}_{\mathbf{q}, \kappa} \label{eq-num-F}
\end{equation}
in terms of the corresponding Berry connection
\begin{equation}
\mathbf{A}_{\mathbf{q}, \kappa}^{\phantom{*}} = i w_{\mathbf{q},
\kappa}^{*} \cdot \nabla_{\mathbf{q}}^{\phantom{*}} w_{\mathbf{q},
\kappa}^{\phantom{*}}. \label{eq-num-A}
\end{equation}
We note that the cross product of two vectors is a scalar in two
dimensions. Also, since the integral in Eq.~(\ref{eq-num-C-1}) is
defined over the conventional Brillouin zone in Fig.~\ref{fig-4}(a),
it must be multiplied by $2$ when calculated over the compact
Brillouin zone in Fig.~\ref{fig-4}(b). Using
Eqs.~(\ref{eq-num-C-1})-(\ref{eq-num-A}), the Majorana Chern number
can then be numerically computed in each flux sector via
\begin{equation}
C = \sum_{\kappa = 1}^{n} C_{\kappa}, \label{eq-num-C-2}
\end{equation}
where the bands $\kappa = 1, \ldots, 2n$ are arranged by increasing
energy eigenvalues so that the summation is over all negative-energy
Majorana bands.

\begin{figure*}[tbp]
\centering
\includegraphics[width=2.0\columnwidth]{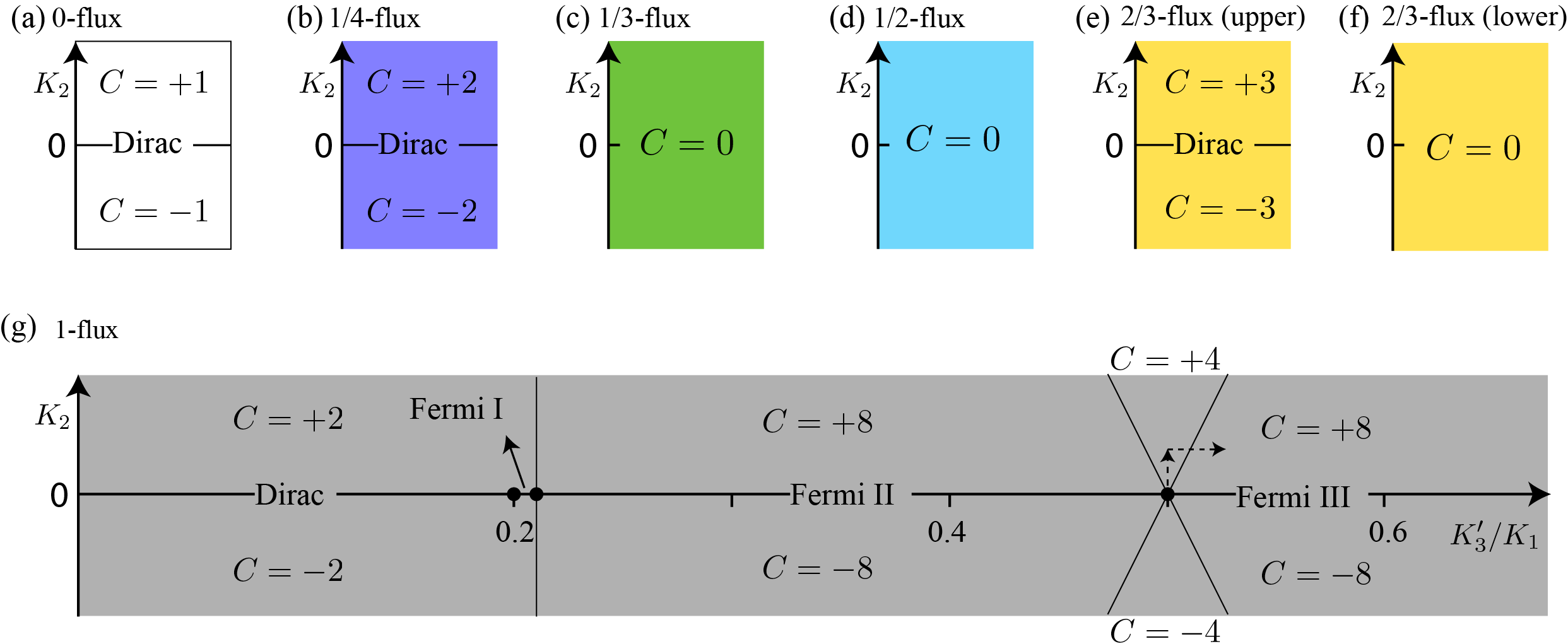}
\caption{(a)-(f) Topological phase diagrams for the various flux
sectors in Fig.~\ref{fig-2} as a function of an infinitesimal $K_2$
at fixed representative values of $K_3$ and $K_3'$. Gapped phases
are labeled by their Majorana Chern numbers $C$, while gapless
phases are labeled by their Majorana nodal structures: Dirac nodes
or Fermi surfaces (i.e., line nodes). (g) Topological phase diagram
for the $1$-flux sector as a function of $K_3' / K_1$ and an
infinitesimal $K_2$. The critical points of the
time-reversal-symmetric model at $K_2 = 0$ are marked by black
circles, while the dashed arrows next to the multicritical point at
$K_2 = 0$ and $K_3' = \frac{1}{2} K_1$ illustrate the two-step
scheme for obtaining the Majorana Chern numbers of the gapped phases
surrounding the multicritical point.} \label{fig-5}
\end{figure*}

The numerical results for the Majorana Chern numbers are summarized
in Fig.~\ref{fig-5}. For most of the flux sectors in
Fig.~\ref{fig-2}, the Chern number only depends on $K_2$ and is
otherwise the same throughout the entire flux sector. In contrast,
for the $2/3$-flux sector, there are two disconnected (``upper'' and
``lower'') phases with distinct Chern numbers, while for the
$1$-flux sector, there are several phases with distinct Chern
numbers that are separated by topological transitions as a function
of $K_3' / K_1$. We note that $K_3 / K_1$ is an irrelevant parameter
in the $1$-flux sector as symmetry-related pairs of $K_3$
interactions [see Fig.~\ref{fig-1}(c)] give rise to equivalent
Majorana hopping terms with a perfect destructive interference
between them \cite{Zhang-2019}.

\subsection{Analytical understanding} \label{sec-ana}

By studying the phase transitions between the various phases in
Fig.~\ref{fig-5}, we can also understand their Majorana Chern
numbers analytically. If the Majorana spectrum is gapped for $K_2 =
0$, the Chern number $C$ vanishes due to time-reversal symmetry and
is robust against an infinitesimally small $K_2 \neq 0$. Thus, the
$1/3$-flux phase, the $1/2$-flux phase, and the ``lower'' $2/3$-flux
phase of Fig.~\ref{fig-2} are all characterized by $C = 0$. If the
Majorana spectrum has gapless nodes for $K_2 = 0$, these nodes are
all gapped out by an infinitesimally small $K_2 \neq 0$, and the
resulting gapped phases have opposite Chern numbers $\pm C$ for
opposite signs of $K_2$. Since a change in the Chern number is
always connected to a closing gap, the Chern number $C$ at $K_2 > 0$
can be understood as a sum of contributions from the various nodes
at $K_2 = 0$.

\begin{figure}[tbp]
\centering
\includegraphics[width=1.0\columnwidth]{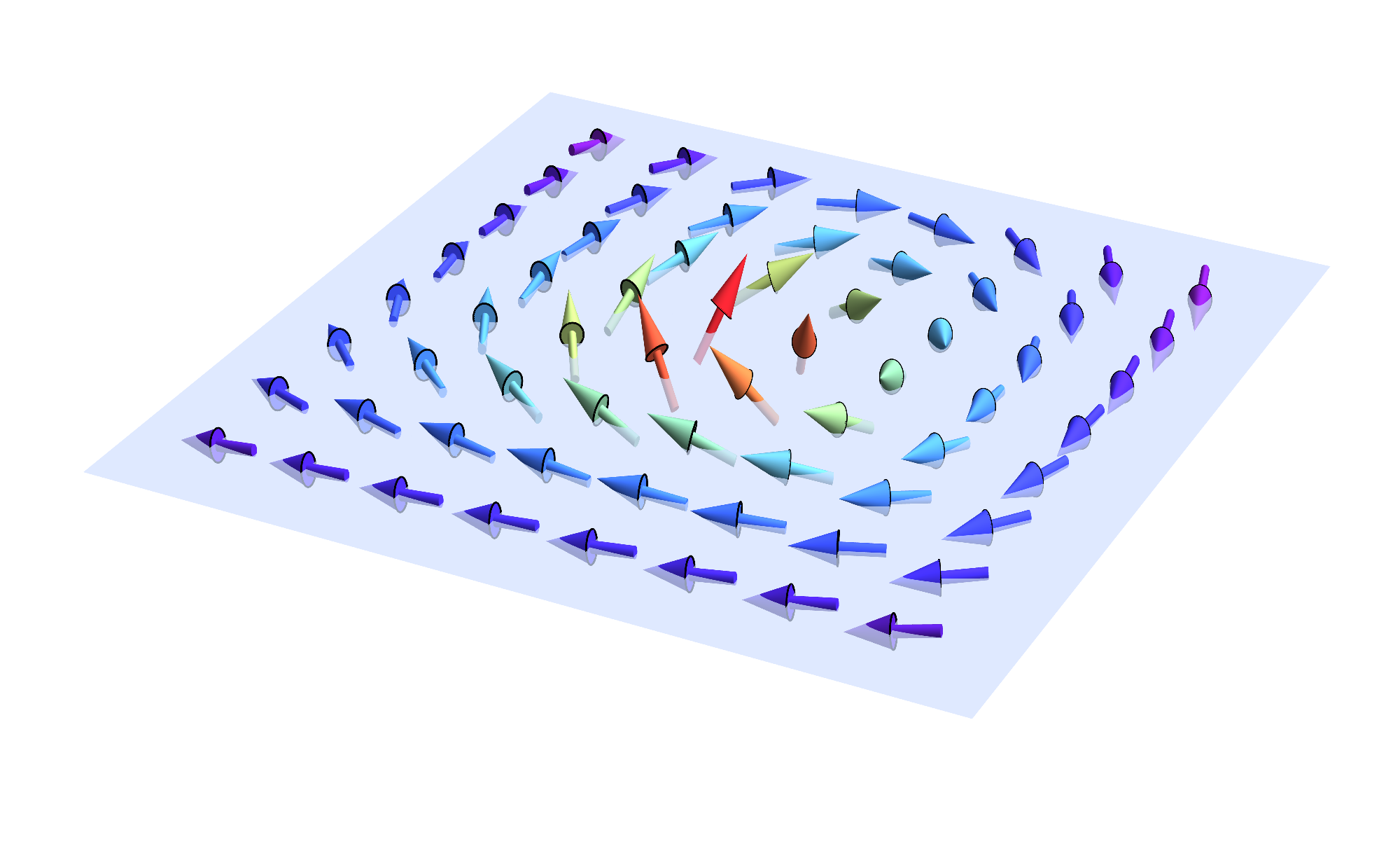}
\caption{Half-skyrmion configuration of the vector field $\mathbf{d}
(\mathbf{q})$ around a Dirac node of winding number $W_{\mathbf{Q}}
= +1$ that is gapped out by a fermion mass $m_{\mathbf{Q}} K_2 >
0$.} \label{fig-6}
\end{figure}

\subsubsection{Dirac nodes} \label{sec-dir}

\begin{figure}[tbp]
\centering
\includegraphics[width=1.0\columnwidth]{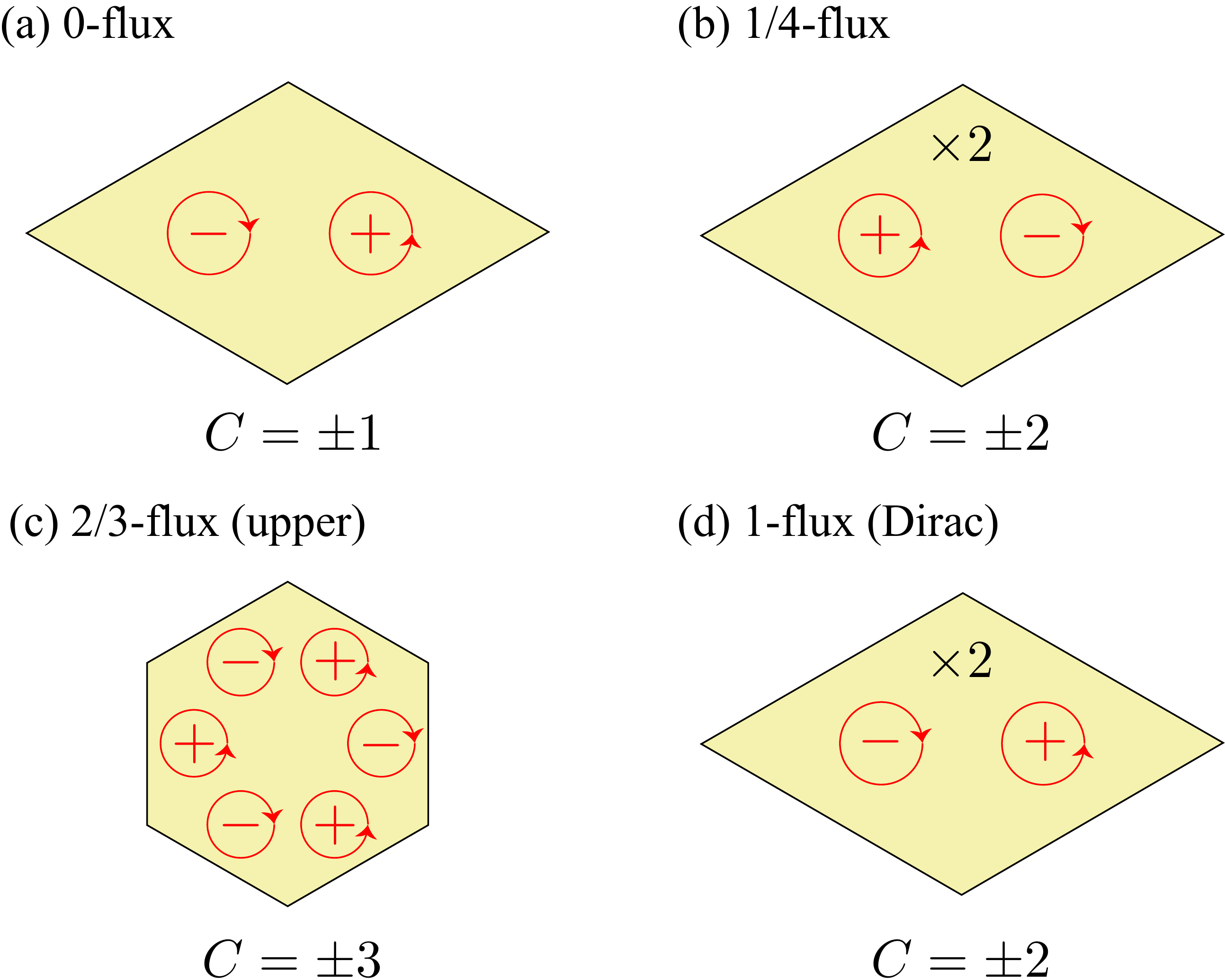}
\caption{Dirac nodes within the compact Brillouin zone in the Dirac
phases of Fig.~\ref{fig-2}. For each Dirac node at momentum
$\mathbf{Q}$, a winding number $W_{\mathbf{Q}}$ of $+1$ ($-1$) is
marked by an anticlockwise (clockwise) arrow, while a positive
(negative) mass coefficient $m_{\mathbf{Q}}$ is marked by a ``$+$''
(``$-$'') label. For $K_2 > 0$, the corresponding contribution to
the Majorana Chern number is either $+1/2$ (red) or $-1/2$ (blue).
If translation symmetry acts projectively on the Majorana fermions,
each contribution must be doubled (``$\times 2$'') due to
translation degeneracy. Summing these contributions, the total
Majorana Chern number $C$ is given by the upper (lower) sign for
$K_2 > 0$ ($K_2 < 0$).} \label{fig-7}
\end{figure}

The low-energy theory around a Dirac node at $K_2 = 0$ and momentum
$\mathbf{Q}$ takes the general form [see Eq.~(\ref{eq-nod-H-2})]
\begin{equation}
\hat{H}_{\mathbf{q}} = \boldsymbol{\beta} (\mathbf{q}) \cdot
\boldsymbol{\tau}, \label{eq-dir-H}
\end{equation}
where $\boldsymbol{\tau} \equiv (\tau_1, \tau_2, \tau_3)$ and, up to
linear order in both $K_2$ and $\delta \mathbf{q} \equiv \mathbf{q}
- \mathbf{Q}$ [see Eqs.~(\ref{eq-nod-beta}) and (\ref{eq-nod-H-3})],
\begin{equation}
\boldsymbol{\beta} (\mathbf{q}) = \left( \boldsymbol{\gamma}_1 \cdot
\delta \mathbf{q}, \, \boldsymbol{\gamma}_2 \cdot \delta \mathbf{q},
\, m_{\mathbf{Q}} K_2 \right). \label{eq-dir-beta}
\end{equation}
For $K_2 \neq 0$, the contribution to the Chern number from the
given Dirac node, $\hat{C}_{\mathbf{Q}}$, is the Chern number of the
negative-energy band in the low-energy theory. For the Hamiltonian
matrix in Eq.~(\ref{eq-dir-H}), this quantity can be calculated by
means of a standard formula \cite{Kitaev-2006}:
\begin{equation}
\hat{C}_{\mathbf{Q}} = \frac{1} {4\pi} \int d \mathbf{q} \,
\mathbf{d} (\mathbf{q}) \cdot \left[ \partial_{q_x} \mathbf{d}
(\mathbf{q}) \times \partial_{q_y} \mathbf{d} (\mathbf{q}) \right],
\label{eq-dir-C-1}
\end{equation}
where $\mathbf{d} (\mathbf{q}) \equiv \boldsymbol{\beta}
(\mathbf{q}) / |\boldsymbol{\beta} (\mathbf{q})|$. Geometrically,
$\hat{C}_{\mathbf{Q}}$ is simply the number of ``skyrmions'' in the
vector field $\mathbf{d} (\mathbf{q})$. Since the vector-field
configuration in Eq.~(\ref{eq-dir-beta}) corresponds to a half
skyrmion or meron (see Fig.~\ref{fig-6}), the contribution of the
given Dirac node to the Chern number becomes
\begin{equation}
\hat{C}_{\mathbf{Q}} = \frac{1}{2} \, W_{\mathbf{Q}} \, \mathrm{sgn}
\left( m_{\mathbf{Q}} K_2 \right) = \pm \frac{1}{2}
\label{eq-dir-C-2}
\end{equation}
in terms of its winding number $W_{\mathbf{Q}}$ [see
Eq.~(\ref{eq-nod-W})]. These contributions of the individual Dirac
nodes are illustrated in Fig.~\ref{fig-7} for the $0$-flux phase,
the $1/4$-flux phase, the ``upper'' $2/3$-flux phase, and the Dirac
$1$-flux phase of Fig.~\ref{fig-2}. For $K_2 > 0$, the resulting
total Chern numbers, $C = \sum_{\mathbf{Q}} \hat{C}_{\mathbf{Q}}$,
are $1$, $2$, $3$, and $2$, respectively.

In general, Dirac nodes emerge in pairs related by inversion
symmetry. Since the two nodes in any pair have opposite mass
coefficients $m_{\mathbf{Q}}$ as well as opposite winding numbers
$W_{\mathbf{Q}}$, each pair contributes $\pm 1$ to the Chern number.
If translation symmetry acts projectively on the Majorana fermions,
corresponding to an overall $\mathbb{Z}_2$ flux in the physical unit
cell, the total Chern number is then necessarily even as these
contributions $\pm 1$ come in identical pairs due to translation
degeneracy.

\subsubsection{Line nodes} \label{sec-lin}

\begin{figure}[tbp]
\centering
\includegraphics[width=1.0\columnwidth]{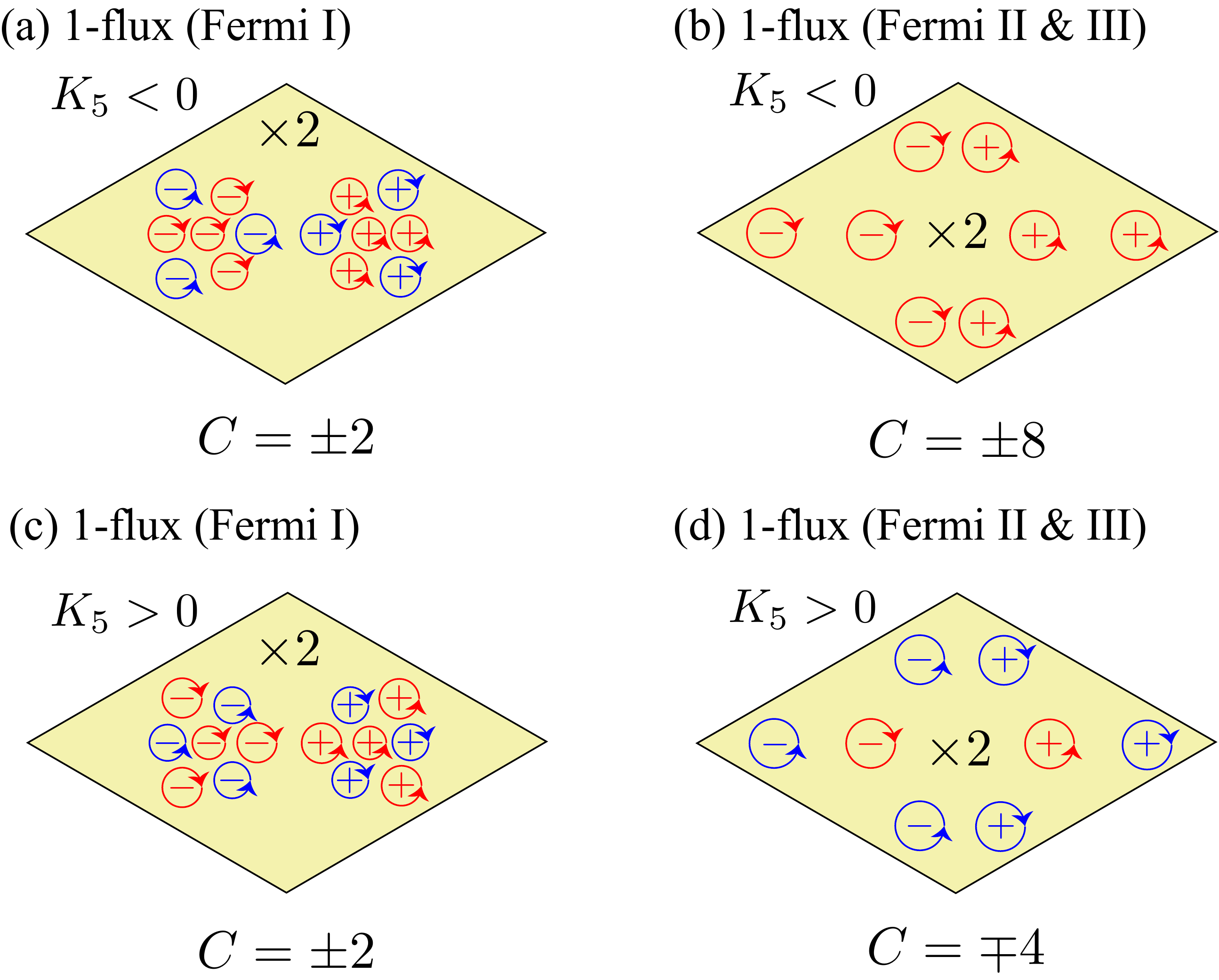}
\caption{Dirac nodes within the compact Brillouin zone in the first
Fermi phase (left) and in the second and the third Fermi phases
(right) of the $1$-flux sector for the two opposite signs (top and
bottom) of a generic fifth-neighbor Majorana hopping amplitude
$K_5$. The notation is identical to Fig.~\ref{fig-7}.} \label{fig-8}
\end{figure}

While the generic Majorana nodal structures at $K_2 = 0$ are point
nodes, these point nodes seem to coexist with line nodes in the
three Fermi $1$-flux phases of Fig.~\ref{fig-2}. As expected, these
accidental line nodes are unstable against generic further-neighbor
Majorana hopping terms that respect the projective symmetries of the
system. In fact, the accidental ``phases'' with line nodes can be
understood as phase transitions between two generic phases with
point nodes as a function of a fifth-neighbor hopping amplitude
$K_5$. While each line node is gapped out into six Dirac nodes at
the same momenta for $K_5 > 0$ and $K_5 < 0$, the winding number of
each Dirac node changes sign at $K_5 = 0$ (see Fig.~\ref{fig-8}).
Thus, Dirac nodes with opposite winding numbers must be connected by
line nodes at the phase transition so that they can exchange their
winding numbers with each other.

Instead of constructing a low-energy theory for each accidental line
node, it is then more natural to include an infinitesimally small
$K_5 \neq 0$ and consider the low-energy theories of the resulting
Dirac nodes. Their contributions to the Chern number at $K_2 \neq
0$, as given by Eq.~(\ref{eq-dir-C-2}), are illustrated in
Fig.~\ref{fig-8} for all three Fermi $1$-flux phases and for both
signs of $K_5$. For the first Fermi phase, the Dirac nodes
corresponding to each line node have a vanishing net contribution to
the Chern number for both $K_5
> 0$ and $K_5 < 0$. Thus, we can deduce that the gapped phase at
$K_5 = 0$ and $K_2 > 0$ is adiabatically connected to that obtained
from the Dirac phase at $K_2 > 0$ and that its total Chern number is
$C = 2$. Conversely, for the second and the third Fermi phases, the
Dirac nodes corresponding to each line node have opposite net
contributions to the Chern number for $K_5 > 0$ and $K_5 < 0$. Thus,
for $K_5 = 0$, there are two possible gapped phases at $K_2 > 0$
with total Chern numbers $8$ and $-4$, respectively. To discriminate
between these two scenarios, we consider the multicritical point at
$K_2 = 0$ and $K_3' = \frac{1}{2} K_1$.

\subsubsection{Multicritical point} \label{sec-mul}

\begin{figure}[tbp]
\centering
\includegraphics[width=0.6\columnwidth]{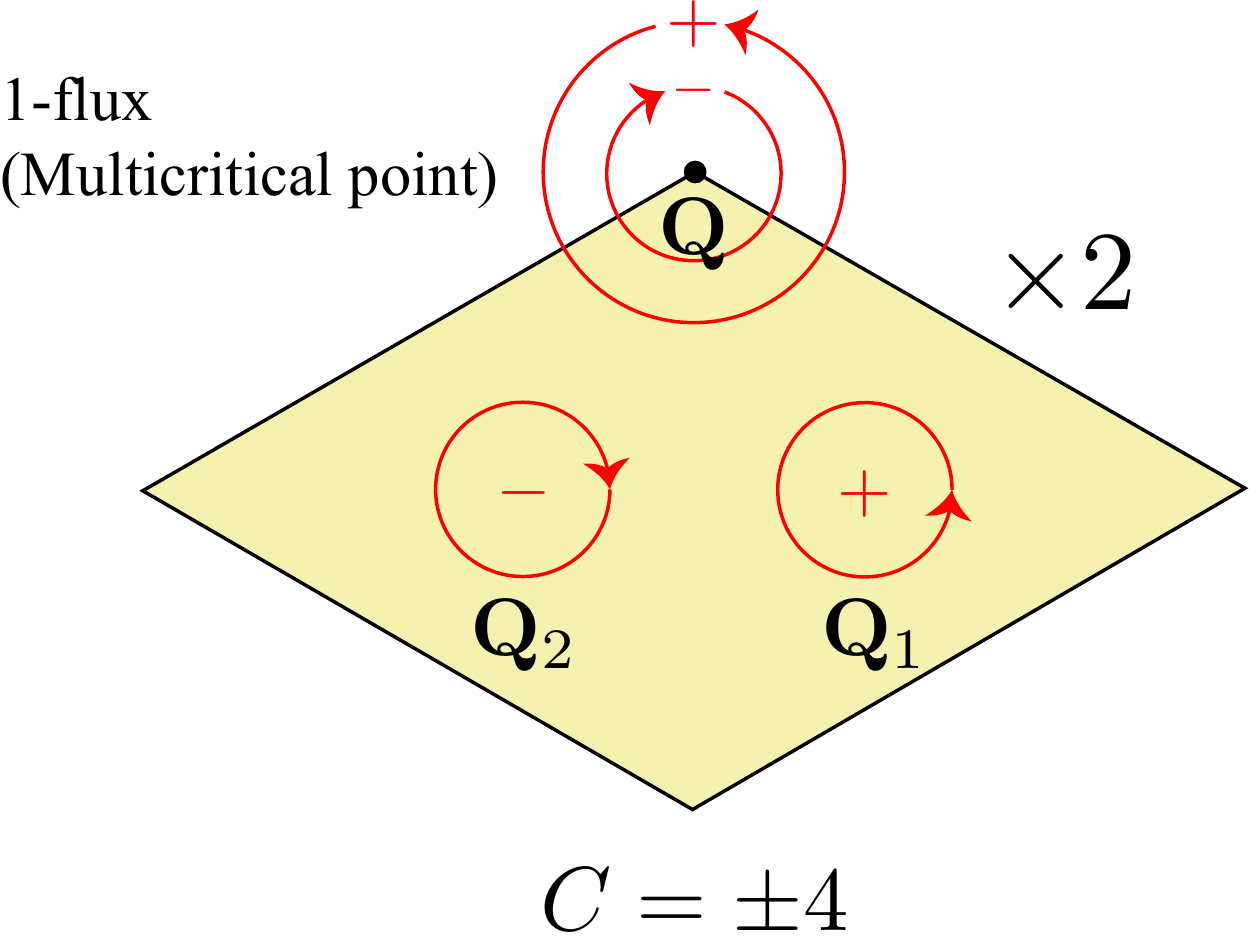}
\caption{Majorana nodes within the compact Brillouin zone at the
multicritical point of the $1$-flux sector: a single Dirac node at
each momentum $\mathbf{Q}_{1,2}$ and a pair of Dirac nodes at
momentum $\mathbf{Q}$. The notation is identical to
Fig.~\ref{fig-7}.} \label{fig-9}
\end{figure}

The second and the third Fermi $1$-flux phases of Fig.~\ref{fig-2}
are separated by a multicritical point [see Fig.~\ref{fig-5}(g)] at
$K_2 = 0$ and $K_3' = \frac{1}{2} K_1$ \cite{footnote-4} where the
accidental line node shrinks to a single momentum $\mathbf{Q}$ (see
Fig.~\ref{fig-9}). Remarkably, at this momentum $\mathbf{Q}$, the $4
\times 4$ Hamiltonian matrix in Eq.~(\ref{eq-ham-H-3}) takes the
\emph{exact} general form
\begin{eqnarray}
H_{\mathbf{Q}} &=& (2 K_3' - K_1) \left( \tau_1 \otimes \eta_0 -
\tau_2 \otimes \eta_1 - \tau_2 \otimes \eta_2 \right)
\nonumber \\
&& + 2 K_2 \left( \tau_0 \otimes \eta_3 - \tau_3 \otimes \eta_1 +
\tau_3 \otimes \eta_2 \right), \label{eq-mul-H-1}
\end{eqnarray}
where ``$\otimes$'' denotes the Kronecker product, while
$\tau_{0,1,2,3}$ and $\eta_{0,1,2,3}$ are the Pauli matrices acting
on the $\mu = A,B$ and $\nu = 1,2$ degrees of freedom, respectively.
Since the two terms in $H_{\mathbf{Q}}$ commute, we can use an
appropriate canonical transformation to recast it in the simpler
form
\begin{equation}
H_{\mathbf{Q}} = \sqrt{3} \, \big[ (2 K_3' - K_1) \tilde{\tau}_0 + 2
K_2 \tilde{\tau}_3 \big] \otimes \tilde{\eta}_3. \label{eq-mul-H-2}
\end{equation}
This matrix has zero eigenvalues, corresponding to phase
transitions, along the lines $K_2 = \pm (K_3' - \frac{1}{2} K_1)$ in
parameter space, while it vanishes identically at the intersection
of these lines, i.e., at the multicritical point. In the following,
we determine the Chern numbers of the gapped phases around the
multicritical point by following the two-step scheme shown in
Fig.~\ref{fig-5}(g).

First, we fix $K_3' = \frac{1}{2} K_1$ and construct the low-energy
theory of the multicritical point node at momentum $\mathbf{Q}$ for
$K_2 \ll K_1$. By expanding $H_{\mathbf{q}}$ in
Eq.~(\ref{eq-ham-H-3}) up to first order in both $K_2$ and $\delta
\mathbf{q} \equiv \mathbf{q} - \mathbf{Q}$, and projecting onto the
basis vectors in Eq.~(\ref{eq-mul-H-2}), we obtain
\begin{eqnarray}
H_{\mathbf{q}} &=& 2 \sqrt{3} \, K_2 \left( \tilde{\tau}_3 \otimes
\tilde{\eta}_3 \right) \label{eq-mul-H-3} \\
&& + \gamma \left[ \delta q_x \left( \tilde{\tau}_1 \otimes
\tilde{\eta}_0 \right) + \delta q_y \left( \tilde{\tau}_2 \otimes
\tilde{\eta}_3 \right) \right]. \nonumber
\end{eqnarray}
Therefore, the low-energy theory contains a pair of Dirac nodes
corresponding to $\tilde{\eta}_3 = \pm 1$. Since the two respective
Dirac nodes have mass coefficients $m_{\mathbf{Q}} = \pm 2 \sqrt{3}$
and winding numbers $W_{\mathbf{Q}} = \pm 1$, they contribute
$\hat{C}_{\mathbf{Q}} = 1$ to the Chern number at $K_2 > 0$.
Together with the contributions $\hat{C}_{\mathbf{Q}_1} =
\hat{C}_{\mathbf{Q}_2} = 1/2$ from the two Dirac nodes at momenta
$\mathbf{Q}_{1,2}$ (see Fig.~\ref{fig-9}), the total Chern number of
the phase at $K_3' = \frac{1}{2} K_1$ and $K_2 > 0$ is then
\begin{equation}
C = 2 \, \big( \hat{C}_{\mathbf{Q}} + \hat{C}_{\mathbf{Q}_1} +
\hat{C}_{\mathbf{Q}_2} \big) = 4, \label{eq-mul-C-1}
\end{equation}
where the additional factor of $2$ comes from translation degeneracy
in the $1$-flux sector.

\begin{figure}[tbp]
\centering
\includegraphics[width=1.0\columnwidth]{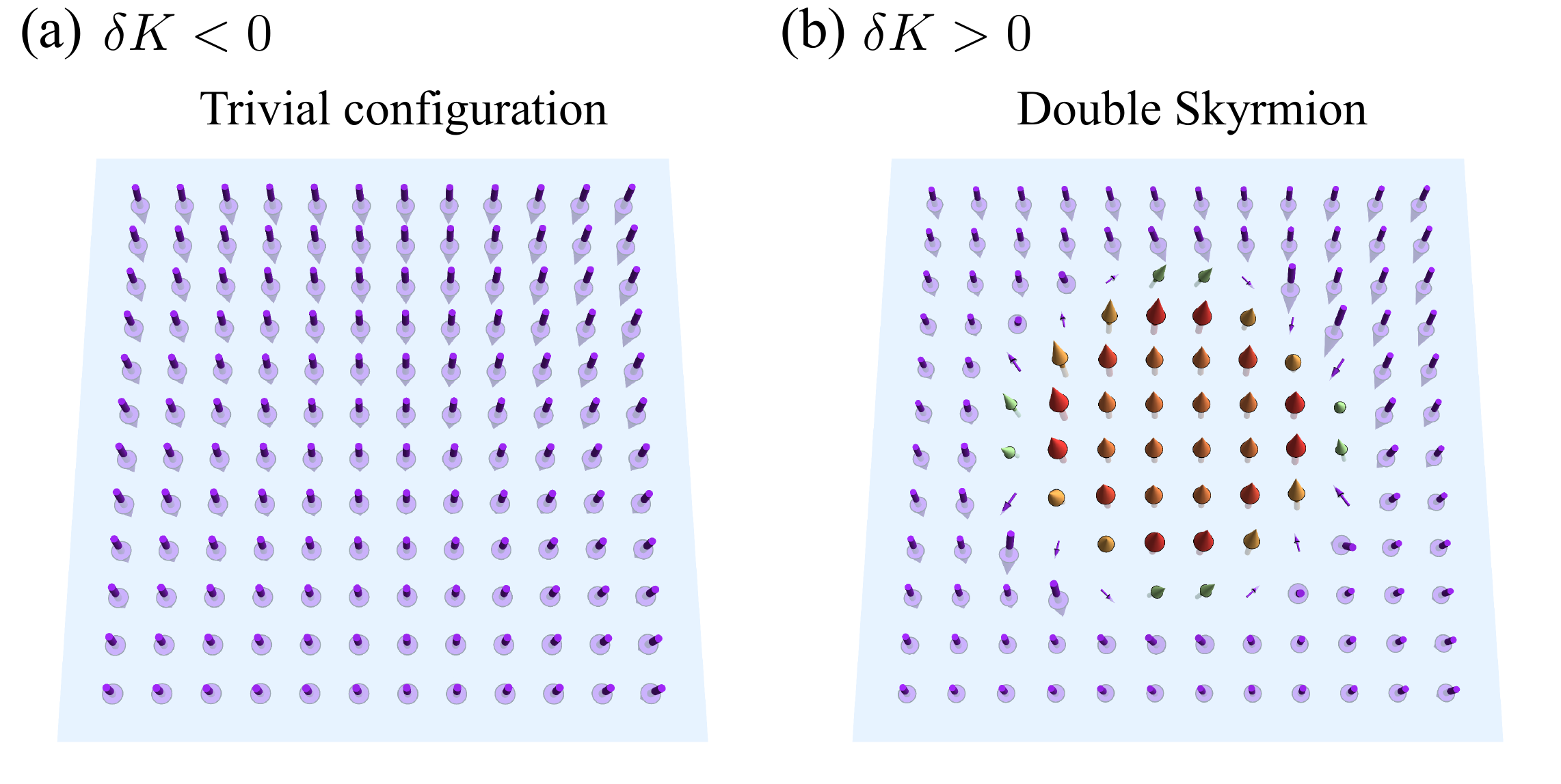}
\caption{Vector-field configuration $\mathbf{d} (\mathbf{q})$ around
the quadratic point node on the two respective sides of the phase
transition, corresponding to (a) $\delta K < 0$ and (b) $\delta K >
0$.} \label{fig-10}
\end{figure}

Next, we fix a particular value of $K_2 > 0$ and consider the phase
transition at $K_3' = \frac{1}{2} K_1 + K_2$. At this phase
transition, the low-energy subspace at the critical momentum
$\mathbf{Q}$ is spanned by the $\tilde{\tau}_3 = -1$ basis vectors
in Eq.~(\ref{eq-mul-H-2}). By expanding $H_{\mathbf{q}}$ in
Eq.~(\ref{eq-ham-H-3}) up to first order in $\delta K \equiv K_3' -
\frac{1}{2} K_1 - K_2$ and second order in $\delta \mathbf{q} \equiv
\mathbf{q} - \mathbf{Q}$, and projecting onto these basis vectors,
we obtain
\begin{eqnarray}
H_{\mathbf{q}} &=& \big[ 2 \sqrt{3} \, \delta K - \theta \big(
\delta q_x^2 + \delta q_y^2 \big) \big] \tilde{\eta}_3
\label{eq-mul-H-4} \\
&& - \chi \left[ \left( \delta q_x^2 - \delta q_y^2 \right)
\tilde{\eta}_1 + 2 \delta q_x \delta q_y \tilde{\eta}_2 \right],
\nonumber
\end{eqnarray}
where $\theta$ and $\chi$ are positive numbers. Therefore, the
low-energy theory of the phase transition is a quadratic point node
at momentum $\mathbf{Q}$. The corresponding change in the Chern
number, $\delta \hat{C}_{\mathbf{Q}}$, across the phase transition
is the difference between the Chern numbers of the negative-energy
bands at $\delta K > 0$ and $\delta K < 0$. Since
Eq.~(\ref{eq-mul-H-4}) assumes the general form of
Eq.~(\ref{eq-dir-H}), these Chern numbers are given by
Eq.~(\ref{eq-dir-C-1}) in terms of the respective vector fields
$\mathbf{d} (\mathbf{q})$ plotted in Fig.~\ref{fig-10}. For $\delta
K < 0$, the vector-field configuration is topologically trivial, and
the Chern number is thus $\hat{C}_{\mathbf{Q}, -} = 0$. For $\delta
K > 0$, the vector-field configuration corresponds to a double
skyrmion, and the Chern number is thus $\hat{C}_{\mathbf{Q}, +} =
2$. Remembering translation degeneracy, the change in the total
Chern number across the phase transition is then
\begin{equation}
\delta C = 2 \, \delta \hat{C}_{\mathbf{Q}} = 2 \, \big(
\hat{C}_{\mathbf{Q}, +} - \hat{C}_{\mathbf{Q}, -} \big) = 4,
\label{eq-mul-C-1}
\end{equation}
and the total Chern number of the gapped phase next to the third
Fermi phase is $C = 4 + 4 = 8$. Since the phase transition at $K_3'
= \frac{1}{2} K_1 - K_2$ is governed by an analogous low-energy
theory, the gapped phase next to the second Fermi phase also has a
total Chern number $C = 8$.

\section{Topological orders} \label{sec-top}

\subsection{Anyon classes and fusion rules} \label{sec-any}

Since the gapped phases described in the previous section are all
topologically ordered, they have anyonic excitations characterized
by particular fusion and braiding properties. Given that gapped
Majorana fermions with a total Chern number $C$ are coupled to a
$\mathbb{Z}_2$ gauge theory, the topological classification is
understood in terms of Kitaev's sixteenfold way \cite{Kitaev-2006}.
While all phases with different Chern numbers $C$ are topologically
distinct theories with different numbers of chiral Majorana edge
modes, and are experimentally distinguishable by their thermal Hall
conductivities, $\kappa_{xy} = \pi C T / 12$, at temperature $T$,
there are only $16$ distinct classes in terms of their bulk anyon
properties, determined by $C \, \mathrm{mod} \, 16$.

\begin{figure}[tbp]
\centering
\includegraphics[width=1.0\columnwidth]{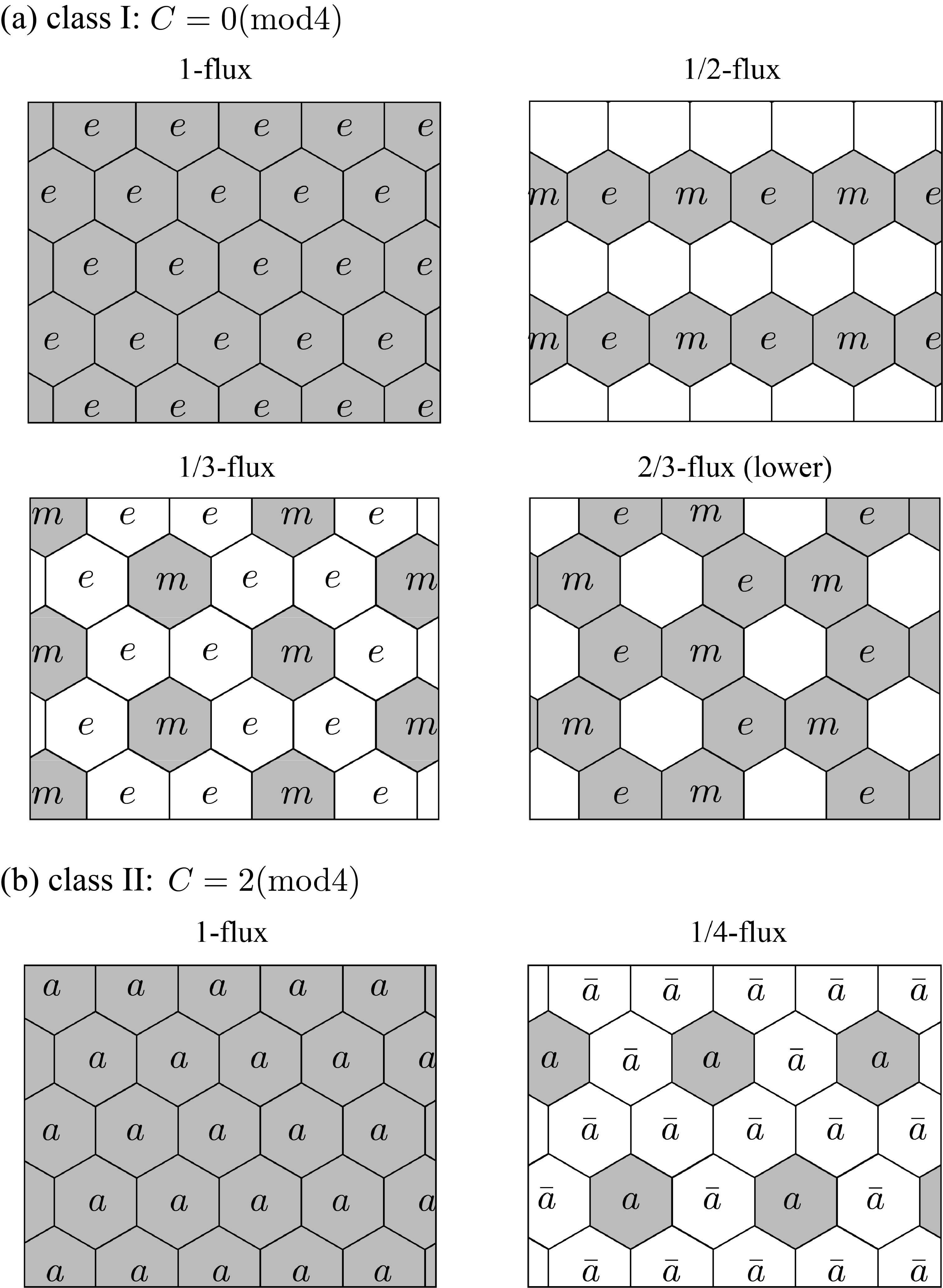}
\caption{Maps of vortex excitations in our topological orders with
even Majorana Chern number $C$. In each case, plaquettes with
ground-state eigenvalues $W_p = +1$ ($W_p = -1$) are marked by white
(gray) filling. At each plaquette, a flux excitation flips the
eigenvalue of $W_p$ and may correspond to vortex classes $e$ or $m$
for $C \, \mathrm{mod} \, 4 = 0$ (a) and vortex classes $a$ or
$\bar{a}$ for $C \, \mathrm{mod} \, 4 = 2$ (b). At each plaquette
with no label, the two classes of vortex excitations are degenerate,
signaling the existence of a ``weak supersymmetry''.} \label{fig-11}
\end{figure}

For each phase, we can use the exact solution of our lattice model
to explicitly identify the topologically distinct anyon classes of
the corresponding theory and to verify the expected fusion rules
between them \cite{Kitaev-2006}. In general, the Majorana fermions
are identified as the fermion excitations $\epsilon$, while the
gauge fluxes are related to the various classes of vortex
excitations. If the Chern number $C$ is odd, there is only one
vortex class $\sigma$, and each flux excitation with respect to the
ground-state flux sector corresponds to such a vortex excitation
$\sigma$. If the Chern number $C$ is even, there are two
topologically distinct vortex classes denoted by $e$ and $m$ for $C
\, \mathrm{mod} \, 4 = 0$ and by $a$ and $\bar{a}$ for $C \,
\mathrm{mod} \, 4 = 2$. Therefore, a flux excitation at any given
plaquette may correspond to either of the two vortex classes. Since
the two vortex classes differ by a fermion in each case, as
indicated by the fusion rules
\begin{eqnarray}
&& \epsilon \times e = m, \quad \,\,\,\, \epsilon \times m = e,
\nonumber \\
&& \epsilon \times a = \bar{a}, \qquad \epsilon \times \bar{a} = a,
\label{eq-any-fus-1}
\end{eqnarray}
the vortex classes corresponding to the various plaquettes can be
mapped out by considering the fermion parity of the ground state
within the flux sector $(p, p')$ that contains two flux excitations
at a general plaquette $p$ and at a far-away reference plaquette
$p'$. If the ground-state fermion parities of the flux sectors
$(p_1, p')$ and $(p_2, p')$ are identical (opposite), the flux
excitations at the plaquettes $p_1$ and $p_2$ correspond to
identical (distinct) vortex excitations. For each phase with even
$C$, the resulting map of the vortex classes is depicted in
Fig.~\ref{fig-11}. We note that the two vortex classes are related
by an anyon permutation symmetry and that the same maps are thus
equally valid with $e \leftrightarrow m$ and $a \leftrightarrow
\bar{a}$.

In terms of the anyon fusion rules, the fermions $\epsilon$ have
similar properties in all phases. Indeed, the general fusion rule
$\epsilon \times \epsilon = 1$ indicates that two fermion
excitations fuse into a topologically trivial excitation. In
contrast, there are three distinct scenarios for the fusion rules
between two identical vortices \cite{Kitaev-2006}. If the Chern
number $C$ is odd, the fusion rule $\sigma \times \sigma = 1 +
\epsilon$ indicates that the vortices are non-Abelian anyons as a
pair of them has a degenerate internal space. The two states in this
internal space correspond to two fusion channels into a trivial
excitation and a fermion excitation, respectively. In the lattice
model, the internal space manifests as a zero-energy fermion in any
flux sector containing two flux excitations far away from each
other. Conversely, if the Chern number $C$ is even, the vortices are
Abelian anyons with only one fusion channel. If $C \, \mathrm{mod}
\, 4 = 0$, the fusion rules
\begin{equation}
e \times e = m \times m = 1 \label{eq-any-fus-2}
\end{equation}
indicate that two identical vortices fuse into a trivial excitation,
whereas if $C \, \mathrm{mod} \, 4 = 2$, the fusion rules
\begin{equation}
a \times a = \bar{a} \times \bar{a} = \epsilon \label{eq-any-fus-3}
\end{equation}
indicate that two identical vortices fuse into a fermion excitation.
In the lattice model, these fusion rules are reflected in the
ground-state fermion parity of a flux sector containing two far-away
flux excitations that correspond to the same vortex class. For $C \,
\mathrm{mod} \, 4 = 0$ ($C \, \mathrm{mod} \, 4 = 2$), this fermion
parity is even (odd) with respect to the overall ground state of the
model.

We finally note that the anyon braiding rules are even more specific
to the topological order than the anyon fusion rules. To explicitly
check these braiding rules in our lattice model, we would need to
calculate the complex hopping matrix elements as one flux excitation
is moved around another one in such a way that the distance between
the two flux excitations is much larger than the correlation length
at each step. Unfortunately, for most of our phases, the correlation
length exceeds the maximal system size for which such a calculation
would be feasible at all. Nevertheless, for all phases with
sufficiently small correlation lengths, the results of such a
calculation are in agreement with the braiding rules in
Ref.~\cite{Kitaev-2006}.

\subsection{Weak symmetry breaking and supersymmetry} \label{sec-wea}

For each map of vortex classes in Fig.~\ref{fig-11}, it is
instructive to consider the interplay between the unbroken lattice
symmetries in the given flux sector and the relevant anyon
permutation symmetry ($e \leftrightarrow m$ or $a \leftrightarrow
\bar{a}$). Interestingly, for the $1/2$-flux phase and the ``lower''
$2/3$-flux phase of Fig.~\ref{fig-2}, certain lattice symmetries
become intertwined with the anyon permutation symmetry $e
\leftrightarrow m$ in the sense that they map plaquettes
corresponding to the two vortex classes $e$ and $m$ onto each other.
In each case, one such lattice symmetry is a twofold rotation around
a $z$ bond separating two $W_p = -1$ plaquettes.

This kind of interplay, commonly known as weak symmetry breaking
\cite{Kitaev-2006}, was first discussed for the spatially
anisotropic gapped phase (``A phase'') of the original Kitaev model,
where the anyon permutation symmetry $e \leftrightarrow m$ is
intertwined with translation symmetry. While there is no symmetry
breaking in the conventional sense as all ground-state correlations
are fully symmetric, there is a symmetry breaking in the topological
properties of the anyonic excitations as symmetries map
topologically distinct anyons onto each other.

Even more remarkably, for both the $1/2$-flux phase and the
``lower'' $2/3$-flux phase of Fig.~\ref{fig-2}, there are certain
plaquettes that do not correspond to any particular vortex class $e$
or $m$ (see Fig.~\ref{fig-11}). Since these plaquettes are mapped
onto themselves by a lattice symmetry that is intertwined with the
anyon permutation symmetry $e \leftrightarrow m$, a vortex
excitation at such a plaquette has a degenerate internal space
consisting of two states that correspond to the two vortex classes
$e$ and $m$. In the lattice model, this degenerate internal space
manifests as a zero-energy fermion in any flux sector that contains
a flux excitation at such a plaquette. The creation and annihilation
operators of the zero-energy fermion can then be identified as the
generators of a fermionic symmetry that is reminiscent of
supersymmetry \cite{Hsieh-2016}.

We emphasize that this fermionic symmetry is not a supersymmetry in
the conventional sense because it relates the vortex excitations $e$
and $m$ that are both bosons in terms of their self statistics
\cite{Kitaev-2006}. Nevertheless, it may be interpreted as a
generalized ``weak supersymmetry'' because it relates topologically
distinct anyonic excitations that are different in their mutual
statistics with respect to each other. We also note that such a
fermionic symmetry does not manifest in the A phase of the original
Kitaev model because translation symmetry does not have a fixed
point and does not map any plaquette onto itself. However, we expect
it to be a generic feature of symmetry-enriched topological order
whenever an anyon permutation symmetry is intertwined with a
point-group symmetry, such as a rotation or a reflection.

\section{Discussion} \label{sec-dis}

In this work, we have realized a wide range of distinct topological
orders in an exactly solvable spin model on the honeycomb lattice.
Each of these topological orders is a $\mathbb{Z}_2$ gauge theory
coupled to Majorana fermions with a total Chern number $C$. Given
their respective Majorana Chern numbers $0$, $\pm 1$, $\pm 2$, $\pm
3$, $\pm 4$, and $\pm 8$, these topological orders correspond to
more than half of Kitaev's sixteenfold way. In particular, the $C =
\pm 3$ phases realize non-Abelian topological orders that are
distinct from the $C = \pm 1$ phases of the Kitaev honeycomb model
both in the number of Majorana edge modes and in the braiding
properties of the non-Abelian anyons. Also, the $C = \pm 8$ and $C =
\pm 4$ phases realize Abelian topological orders in which the gauge
fluxes have fermionic and semionic particle statistics, respectively
\cite{Kitaev-2006}. Since our model emerges naturally from the
Kitaev honeycomb model in the presence of perturbations
\cite{Zhang-2019}, we expect that its topological orders are likely
to be realized in spin-orbit-coupled honeycomb magnets, such as
$\alpha$-RuCl$_3$.

From an experimental perspective, the key signature of each $C \neq
0$ topological order is a specific quantized value of the thermal
Hall conductivity, $\kappa_{xy} = \pi C T / 12$, at any temperature
$T$ below the bulk energy gap. This quantized value is directly
proportional to the chiral central charge, $c = C/2$, of the edge
theory, whose integer (fractional) values correspond to Abelian
(non-Abelian) bulk topological orders. In the fractional-flux
sectors, the topological order is also accompanied by a spontaneous
breaking of translation symmetry which, in the presence of any
spin-lattice coupling, gives rise to a periodic lattice distortion
and can thus be picked up with nuclear magnetic resonance or elastic
x-ray scattering. Moreover, the spontaneous breaking of discrete
translation leads to a finite-temperature phase transition
\cite{Zhang-2019} that is readily observable in the specific heat.

Conceptually, the generalization of the Kitaev honeycomb model in
this work facilitates a convenient band-structure engineering for
Majorana fermions coupled to a $\mathbb{Z}_2$ gauge theory. In many
ways, the resulting topological phases are analogous to those
studied in the context of noninteracting electrons. For example, the
$C \neq 0$ topological orders in this work correspond to Chern
insulators as the Majorana fermions form topologically nontrivial
bands with finite Chern numbers. In the future, it would be
interesting to engineer other topological band structures for the
Majorana fermions and thereby realize Majorana analogs of other
noninteracting topological phases, such as topological insulators.
Alternatively, it could be worth completing the realization of
Kitaev's sixteenfold way by engineering Majorana band structures
with total Chern numbers $\pm 5$, $\pm 6$, and $\pm 7$.

We finally note that, compared to the other flux sectors in this
work, the Majorana fermions have a different behavior in the
$3/4$-flux sector because inversion symmetry acts differently on
them. In the presence of time-reversal symmetry, the entire Majorana
spectrum is twofold degenerate, and the Dirac nodes thus come in
pairs at each nodal momentum. If time-reversal symmetry is broken,
these Dirac nodes then expand into line nodes instead of gapping
out. While the line nodes are stable in the noninteracting Majorana
theory, they are expected to have instabilities against Majorana
interactions \cite{Hermanns-2015}. The study of these instabilities
and the resulting topological phases is the subject of ongoing
further work.

\begin{acknowledgments}

We thank Yong Baek Kim for useful discussions. S.-S.~Z. and
C.~D.~B.~are supported by funding from the Lincoln Chair of
Excellence in Physics. The work of G.~B.~H.~at ORNL was supported by
Laboratory Director's Research and Development funds.

\end{acknowledgments}


\begin{references}

\bibitem{Wen-2004} X.-G. Wen, \emph{Quantum Field Theory of Many-Body
Systems} (Oxford University Press, Oxford, 2004).
\bibitem{Chen-2010} X. Chen, Z.-C. Gu, and X.-G. Wen, Phys. Rev. B
\textbf{82}, 155138 (2010).
\bibitem{Wen-1990} X. G. Wen and Q. Niu, Phys. Rev. B \textbf{41},
9377 (1990).
\bibitem{Preskill-2006} A. Kitaev and J. Preskill, Phys. Rev. Lett.
\textbf{96}, 110404 (2006).
\bibitem{Levin-2006} M. Levin and X.-G. Wen, Phys. Rev. Lett.
\textbf{96}, 110405 (2006).
\bibitem{Kitaev-2003} A. Y. Kitaev, Ann. Phys. \textbf{303}, 2 (2003).
\bibitem{Nayak-2008} C. Nayak, S. H. Simon, A. Stern, M. Freedman, and
S. Das Sarma, Rev. Mod. Phys. \textbf{80}, 1083 (2008).
\bibitem{Kitaev-2006} A. Y. Kitaev, Ann. Phys. \textbf{321}, 2 (2006).
\bibitem{Mermin-1979} N. D. Mermin, Rev. Mod. Phys. \textbf{51}, 591
(1979).
\bibitem{Bernevig-2015} A. Bernevig and T. Neupert, arXiv:1506.05805.
\bibitem{Kasahara-2018} Y. Kasahara, T. Ohnishi, Y. Mizukami, O.
Tanaka, S. Ma, K. Sugii, N. Kurita, H. Tanaka, J. Nasu, Y. Motome,
T. Shibauchi, and Y. Matsuda, Nature \textbf{559}, 227 (2018).
\bibitem{Jackeli-2009} G. Jackeli and G. Khaliullin, Phys. Rev. Lett.
\textbf{102}, 017205 (2009).
\bibitem{Chaloupka-2010} J. Chaloupka, G. Jackeli, and G.
Khaliullin, Phys. Rev. Lett. \textbf{105}, 027204 (2010).
\bibitem{Liu-2018} H. Liu and G. Khaliullin, Phys. Rev. B \textbf{97},
014407 (2018).
\bibitem{Sano-2018} R. Sano, Y. Kato, and Y. Motome, Phys. Rev. B
\textbf{97}, 014408 (2018).
\bibitem{Li-2017} F.-Y. Li, Y.-D. Li, Y. Yu, A. Paramekanti, and G.
Chen, Phys. Rev. B \textbf{95}, 085132 (2017).
\bibitem{Jang-2019} S.-H. Jang, R. Sano, Y. Kato, and Y. Motome,
Phys. Rev. B \textbf{99}, 241106(R) (2019).
\bibitem{Singh-2010} Y. Singh and P. Gegenwart, Phys. Rev. B
\textbf{82}, 064412 (2010).
\bibitem{Liu-2011} X. Liu, T. Berlijn, W.-G. Yin, W. Ku, A. Tsvelik,
Y.-J. Kim, H. Gretarsson, Y. Singh, P. Gegenwart, and J. P. Hill,
Phys. Rev. B \textbf{83}, 220403(R) (2011).
\bibitem{Singh-2012} Y. Singh, S. Manni, J. Reuther, T. Berlijn, R.
Thomale, W. Ku, S. Trebst, and P. Gegenwart, Phys. Rev. Lett.
\textbf{108}, 127203 (2012).
\bibitem{Choi-2012} S. K. Choi, R. Coldea, A. N. Kolmogorov, T.
Lancaster, I. I. Mazin, S. J. Blundell, P. G. Radaelli, Y. Singh, P.
Gegenwart, K. R. Choi, S.-W. Cheong, P. J. Baker, C. Stock, and J.
Taylor, Phys. Rev. Lett. \textbf{108}, 127204 (2012).
\bibitem{Ye-2012} F. Ye, S. Chi, H. Cao, B. C. Chakoumakos, J. A.
Fernandez-Baca, R. Custelcean, T. F. Qi, O. B. Korneta, and G. Cao,
Phys. Rev. B \textbf{85}, 180403(R) (2012).
\bibitem{Comin-2012} R. Comin, G. Levy, B. Ludbrook, Z.-H. Zhu, C. N.
Veenstra, J. A. Rosen, Y. Singh, P. Gegenwart, D. Stricker, J. N.
Hancock, D. van der Marel, I. S. Elfimov, and A. Damascelli, Phys.
Rev. Lett. \textbf{109}, 266406 (2012).
\bibitem{Chun-2015} S. Hwan Chun, J.-W. Kim, J. Kim, H. Zheng, C. C.
Stoumpos, C. D. Malliakas, J. F. Mitchell, K. Mehlawat, Y. Singh, Y.
Choi, T. Gog, A. Al-Zein, M. M. Sala, M. Krisch, J. Chaloupka, G.
Jackeli, G. Khaliullin, and B. J. Kim, Nat. Phys. \textbf{11}, 462
(2015).
\bibitem{Williams-2016} S. C. Williams, R. D. Johnson, F. Freund, S.
Choi, A. Jesche, I. Kimchi, S. Manni, A. Bombardi, P. Manuel, P.
Gegenwart, and R. Coldea, Phys. Rev. B \textbf{93}, 195158 (2016).
\bibitem{Kitagawa-2018} K. Kitagawa, T. Takayama, Y. Matsumoto, A.
Kato, R. Takano, Y. Kishimoto, R. Dinnebier, G. Jackeli, and H.
Takagi, Nature \textbf{554}, 341 (2018).
\bibitem{Plumb-2014} K. W. Plumb, J. P. Clancy, L. J. Sandilands, V.
V. Shankar, Y. F. Hu, K. S. Burch, H.-Y. Kee, and Y.-J. Kim, Phys.
Rev. B \textbf{90}, 041112(R) (2014).
\bibitem{Sandilands-2015} L. J. Sandilands, Y. Tian, K. W. Plumb,
Y.-J. Kim, and K. S. Burch, Phys. Rev. Lett. \textbf{114}, 147201
(2015).
\bibitem{Sears-2015} J. A. Sears, M. Songvilay, K. W. Plumb, J. P.
Clancy, Y. Qiu, Y. Zhao, D. Parshall, and Y.-J. Kim, Phys. Rev. B
\textbf{91}, 144420 (2015).
\bibitem{Majumder-2015} M. Majumder, M. Schmidt, H. Rosner, A. A.
Tsirlin, H. Yasuoka, and M. Baenitz, Phys. Rev. B \textbf{91},
180401(R) (2015).
\bibitem{Johnson-2015} R. D. Johnson, S. C. Williams,
A. A. Haghighirad, J. Singleton, V. Zapf, P. Manuel, I. I. Mazin, Y.
Li, H. O. Jeschke, R. Valent\'i, and R. Coldea, Phys. Rev. B
\textbf{92}, 235119 (2015).
\bibitem{Sandilands-2016} L. J. Sandilands, Y. Tian, A. A. Reijnders,
H.-S. Kim, K. W. Plumb, Y.-J. Kim, H.-Y. Kee, and K. S. Burch, Phys.
Rev. B \textbf{93}, 075144 (2016).
\bibitem{Banerjee-2016} A. Banerjee, C. A. Bridges, J.-Q. Yan, A. A.
Aczel, L. Li, M. B. Stone, G. E. Granroth, M. D. Lumsden, Y. Yiu, J.
Knolle, S. Bhattacharjee, D. L. Kovrizhin, R. Moessner, D. A.
Tennant, D. G. Mandrus, and S. E. Nagler, Nat. Mater. \textbf{15},
733 (2016).
\bibitem{Banerjee-2017} A. Banerjee, J. Yan, J. Knolle, C. A. Bridges,
M. B. Stone, M. D. Lumsden, D. G. Mandrus, D. A. Tennant, R.
Moessner, and S. E. Nagler, Science \textbf{356}, 1055 (2017).
\bibitem{Do-2017} S.-H. Do, S.-Y. Park, J. Yoshitake, J. Nasu, Y.
Motome, Y. S. Kwon, D. T. Adroja, D. J. Voneshen, K. Kim, T.- H.
Jang, J.-H. Park, K.-Y. Choi, and S. Ji, Nat. Phys. \textbf{13},
1079 (2017).
\bibitem{Yan-2019} J.-Q. Yan, S. Okamoto, Y. Wu, Q. Zheng, H. D. Zhou,
H. B. Cao, and M. A. McGuire, Phys. Rev. Materials \textbf{3},
074405 (2019).
\bibitem{Xing-2019} J. Xing, H. Cao, E. Emmanouilidou, C. Hu, J. Liu,
D. Graf, A. P. Ramirez, G. Chen, and N. Ni, arXiv:1903.03615.
\bibitem{Sala-2019} G. Sala, M. B. Stone, B. K. Rai, A. F. May, D. S.
Parker, G. B. Hal\'asz, Y. Q. Cheng, G. Ehlers, V. O. Garlea, Q.
Zhang, M. D. Lumsden, A. D. Christianson, arXiv:1907.10627.
\bibitem{Zhang-2019} S.-S. Zhang, Z. Wang, G. B. Hal\'asz, and C. D.
Batista, Phys. Rev. Lett. \textbf{123}, 057201 (2019).
\bibitem{footnote-1} In this work, two sites are ``$r$-th neighbors''
if the shortest path connecting them consists of $r$ bonds.
\bibitem{Lieb-1994} E. H. Lieb, Phys. Rev. Lett. \textbf{73}, 2158
(1994).
\bibitem{Wen-2002} X.-G. Wen, Phys. Rev. B \textbf{65}, 165113
(2002).
\bibitem{You-2012} Y.-Z. You, I. Kimchi, and A. Vishwanath, Phys.
Rev. B \textbf{86}, 085145 (2012).
\bibitem{footnote-2} Since $\det H_{\mathbf{Q}} = \pm | \det
M_{\mathbf{Q}} |^2 = 0$ at the nodal momentum $\mathbf{Q}$, the
matrix $M_{\mathbf{Q}}$ must have a zero eigenvalue.
\bibitem{footnote-3} In principle, non-generic quadratic point nodes
could also appear in the Majorana spectrum. However, since they are
absent in all flux sectors, we restrict our analysis to generic
Dirac nodes.
\bibitem{footnote-4} We note that this multicritical point only
appears to be a ``multicritical line'' in Fig.~\ref{fig-2} because
one of its axes is the irrelevant parameter $K_3 / K_1$.
\bibitem{Hsieh-2016} T. H. Hsieh, G. B. Hal\'asz, and T. Grover,
Phys. Rev. Lett. \textbf{117}, 166802 (2016).
\bibitem{Hermanns-2015} M. Hermanns, S. Trebst, and A. Rosch, Phys.
Rev. Lett. \textbf{115}, 177205 (2015).

\end{references}
\end{document}